\documentclass{emulateapj}

\slugcomment{Accepted for publication in ApJ}

\shorttitle{Multiwavelength Study of PKS~2101--490}
\shortauthors{Godfrey et al.}

\begin{document}

\title{A Multi-Wavelength Study of the Jet, Lobes and Core of the Quasar PKS~2101--490}

\author{L.E.H. Godfrey\altaffilmark{1,2}, G.V. Bicknell\altaffilmark{2},
  J.E.J. Lovell\altaffilmark{3,4,5}, D.L. Jauncey\altaffilmark{3},
  J. Gelbord\altaffilmark{6}, D. A. Schwartz\altaffilmark{7}, E.S.~Perlman\altaffilmark{8},
  H. L. Marshall\altaffilmark{9}, M.~Birkinshaw\altaffilmark{7,10}, D. M. Worrall\altaffilmark{7,10}, M. Georganopoulos\altaffilmark{11}, D.W.Murphy\altaffilmark{12}}

\email{L.Godfrey@curtin.edu.au}

\altaffiltext{1}{International Centre for Radio Astronomy Research, Curtin University, GPO Box U1987, Perth, WA, 6102, Australia}
\altaffiltext{2}{Research School of Astronomy and Astrophysics, Australian National
University, Cotter Road, Weston, ACT, 2611, Australia}
\altaffiltext{3}{Australia Telescope National Facility, CSIRO, P.O. Box 76, Epping, NSW,
2121, Australia}
\altaffiltext{4}{CSIRO, Industrial Physics, PO Box 218 Lindfield NSW 2070, Australia}
\altaffiltext{5}{School of Mathematics and Physics, University of Tasmania, Tas 7001, Australia}
\altaffiltext{6}{Department of Physics, Durham University, South Road, Durham, DH1 3LE, UK}
\altaffiltext{7}{Harvard-Smithsonian Center for Astrophysics, 60 Garden Street, Cambridge, MA 02138, USA}
\altaffiltext{8}{Physics and Space Sciences Department, Florida Institute of Technology, 150 West University Boulevard, Melbourne, FL 32901, USA}
\altaffiltext{9}{Kavli Institute for Astrophysics and Space Research, Massachusetts Institute of Technology, USA}
\altaffiltext{10}{HH Wills Physics Laboratory, University of Bristol, Tyndall Avenue, Bristol BS8 1TL, UK}
\altaffiltext{11}{Department of Physics, Joint Center for Astrophysics, University of Maryland-Baltimore County, 1000 Hilltop Circle, Baltimore, MD 21250, USA}
\altaffiltext{12}{Jet Propulsion Laboratory, 4800 Oak Grove Drive, Pasadena, CA 91109, USA}

\begin{abstract}

We present a detailed study of the X-ray, optical and radio emission from the jet, lobes and core of the quasar PKS~2101--490 as revealed by new \emph{Chandra}, HST and ATCA images. We extract the radio to X-ray spectral energy distributions from seven regions of the 13$\arcsec$ jet, and model the jet X-ray emission in terms of Doppler beamed inverse Compton scattering of the cosmic microwave background (IC/CMB) for a jet in a state of equipartition between particle and magnetic field energy densities. This model implies that the jet remains highly relativistic hundreds of kpc from the nucleus, with a bulk Lorentz factor $\Gamma \sim 6$ and magnetic field of order 30~$\mu$G. We detect an apparent radiative cooling break in the synchrotron spectrum of one of the jet knots, and are able to interpret this in terms of a standard one-zone continuous injection model, based on jet parameters derived from the IC/CMB model. However, we note apparent substructure in the bright optical knot in one of the HST bands. We confront the IC/CMB model with independent estimates of the jet power, and find that the IC/CMB model jet power is consistent with the independent estimates, provided that the minimum electron Lorentz factor $\gamma_{\rm min} \gtrsim 50$, and the knots are significantly longer than the jet width, as implied by de-projection of the observed knot lengths.

\end{abstract}


\section{Introduction} \label{sec:intro}

The first \emph{Chandra} observations of the quasar PKS~0637--752 revealed a bright X-ray jet extending 12$\arcsec$ from the quasar core ($>$ 500 kpc de-projected, assuming a jet viewing angle $\theta < 9^{\circ}$, as evidence by the observed proper motions of the pc-scale jet using a modern cosmology \citep{lovell00}), associated with the previously known radio jet, but with an unexpectedly high X-ray to radio flux density ratio \citep{schwartz00, chartas00}. Since then, tens of quasar jets have been found to possess X-ray jets with similarly high X-ray to radio flux density ratios \citep[e.g.][]{harris02, sambruna02, sambruna04, marshall05, marshall11, kataoka05, massaro11}. The strong X-ray emission from kpc-scale quasar jets such as that of PKS~0637--752 is hard to explain in terms of standard emission mechanisms such as thermal bremsstrahlung or synchrotron self Compton emission \citep{chartas00, schwartz00}. A popular explanation for the strong X-ray emission is the beamed, equipartition IC/CMB model proposed for PKS~0637--752 by \citet{tavecchio00} and \citet{celotti01}, in which the flow velocity is assumed to be highly relativistic and directed close to the line of sight. A relativistic jet velocity increases the energy density of the cosmic microwave background (CMB) in the rest frame of the jet plasma, thereby increasing the X-ray emissivity produced via inverse Compton scattering of CMB photons. The small jet viewing angle implies that the emission is Doppler beamed towards the observer. The appeal of this model is largely due to its simplicity and consistency with equipartition between magnetic and particle energy densities in the emitting plasma. From here on, we refer to the beamed, equipartition IC/CMB model as simply the IC/CMB model. 

A number of uncertainties and potential problems for the IC/CMB model have been identified: (1) There is no conclusive theoretical or empirical justification for the assumption of equipartition in jet plasma, although, for a given jet speed, the equipartition condition minimises the plasma energy density and jet power.  (2) The IC/CMB model requires jet Lorentz factors of order $\Gamma \sim 5 - 25$ on scales of hundreds of kpc from the core. Such large jet Lorentz factors are inconsistent with studies of the radio emission from large-scale jets and counter-jets \citep[$\Gamma \lesssim 1.5$][]{wardle97, mullin09}. A suggested solution to this problem invokes velocity structure across the jet --- the so-called ``spine-sheath" model, in which the radio emission from jets in FRII radio galaxies is dominated by a slower moving sheath, whilst the emission from quasar jets with small viewing angles is said to be dominated by the Doppler boosted radiation from a fast moving spine \citep[see e.g.][]{hardcastle06, mullin09}. (3) It has been argued that, due to the long cooling timescale for the $\gamma \sim 100$ electrons responsible for the IC/CMB X-ray emission, radiative and adiabatic losses alone cannot account for the rapid drop in X-ray surface brightness outside the knots \citep{tavecchio03, stawarz04a, siemiginowska07}. (4) The IC/CMB model makes strong, testable predictions about the redshift dependence of kpc-scale X-ray jets. Specifically, the model predicts that the X-ray surface brightness should be  redshift independent, because the CMB energy density increases as (1+z)$^4$ which balances the usual (1+z)$^{-4}$ decrease of surface brightness with redshift. Therefore, the ratio of X-ray to radio surface brightness should be a strong function of redshift \citep{schwartz02, marshall11}. So far, the search for the predicted redshift dependence has been unsuccessful \citep{marshall11, kataoka05}. (5) It has been argued \citep[e.g.][]{atoyan04} that the very large jet powers derived from the IC/CMB model ($\gtrsim 10^{48}$ ergs/s, e.g.\, \citet{tavecchio00}) are prohibitively large. Such high jet power is disfavoured because $10^{48}$~ergs/s is equal to the Eddington luminosity of a $10^{10}$ M$_\odot$ black hole, and such high jet power is an outlier when compared to samples of FRII radio galaxies such as the \citet{rawlings91} sample, for which the largest estimated jet power is of order $10^{47}$~ergs/s. (6) A number of sources show a decreasing X-ray to radio flux density ratio along the jet, which, if the IC/CMB model is correct, implies deceleration must be taking place on hundreds of kpc-scales \citep{georganopoulos04}. However, it is not clear how the gradual deceleration can occur, and there is no evidence for deceleration in lobe-dominated radio galaxies, which may be expected in such a model \citep{hardcastle06}.

Despite the numerous concerns surrounding the IC/CMB model, none of the issues listed above is currently seen as definitively refuting the model, and it continues to receive attention in the literature as the likely candidate for the quasar jet X-ray emission mechanism. In this paper, we critically assess the application of the IC/CMB model to jet X-ray emission of PKS~2101--490.

PKS 2101-490 was first reported as a bright flat spectrum radio source by \citet{ekers69}. \citet{marshall05} reported a redshift of z $\approx$ 1.04 for this source, based on an unpublished Magellan spectrum (see also the discussion in \S \ref{sec:core}). The spectroscopic redshift determination is robust, with uncertainty of approximately $\pm$0.003. Further details of the spectroscopic observations and data analysis will be presented in an upcoming paper (Gelbord \& Marshall, in prep.). 

Studies at the Australia Telescope Compact Array (ATCA) revealed significant radio emission 
on arcsecond scales \citep{lovell97}. For this reason, PKS 2101--490 was included in the \emph{Chandra} 
snapshot survey of flat spectrum radio quasars with arcsecond scale radio jets \citep{marshall05}. 
\citet{marshall05} presented an 8.6~GHz ATCA image along with a 5 kilosecond snapshot \emph{Chandra} image (\dataset [ADS/Sa.CXO#obs/03126] {Chandra ObsID 3126}) that revealed significant X-ray emission associated with the 13$\arcsec$ eastern radio jet. Based on the results of the snapshot survey, and its morphological similarity to PKS~0637--752, PKS~2101--490 was selected along with a number of other sources: PKS~1421-490 \citep{godfrey09, gelbord05}; PKS~1055+201 \citep{schwartz06b}; PKS~0208-512 \citep{perlman11}; PKS~1202-262 \citep{perlman11}; PKS~0920-397 \citep{schwartz10}; and PKS~1030-357, as a target for deeper follow-up observations with \emph{Chandra}, HST and ATCA. Here we present a detailed physical analysis of the jet, hotspot and lobes of this source based on new ATCA, \emph{Chandra} and HST follow-up images.

\begin{deluxetable}{cccc}
\tabletypesize{\scriptsize}
\tablecaption{Observation Information \label{table:2101_obs_info}}
\tablewidth{0pt}
\tablehead{
\colhead{Instrument} & \colhead{Band} & \colhead{Mode} & \colhead{Date} 
}
\startdata
ATCA & 4.8~GHz & 1.5A/6A & May 25/Sep 2 2000 \\
ATCA & 8.64~GHz & 1.5A/6A & May 25/Sep 2 2000  \\
ATCA & 17.73~GHz & 6C & May 10 2004 \\
ATCA & 20.16~GHz & 6C & May 10 2004  \\
HST & F814W & ACS/WFC & Mar 8 2005  \\
HST & F475W & ACS/WFC & Mar 8 2005  \\
\emph{Chandra} & $0.5-7$~keV & ACIS-S3 & Dec 17 2004 
\enddata
\end{deluxetable}

In \S \ref{sec:observations} we describe the observations and data reduction. In \S \ref{sec:results} we discuss the method and results of modelling the spectral energy distributions of spatially separated jet knots in terms of the IC/CMB model. In \S \ref{sec:jet_power} we compare independent jet power estimates with that obtained from the IC/CMB model of jet X-ray emission. In \S \ref{sec:2101_K6_broken_powerlaw} we discuss the optical emission detected from one of the jet knots, and present an interpretation of this in terms of a broken power law synchrotron spectrum. In \S \ref{sec:core} we discuss the X-ray spectrum of the quasar core. In \S \ref{sec:lobes} we discuss the radio and X-ray emission from the lobes, in particular, we model the lobe SEDs in terms of inverse Compton scattering of the CMB. In \S \ref{sec:conclusions} we present the conclusions and final remarks. 

\section{Observations and Data Reduction} \label{sec:observations}

\subsection{Overview}
\begin{figure*}[!ht]
\epsscale{1.05}
\plottwo{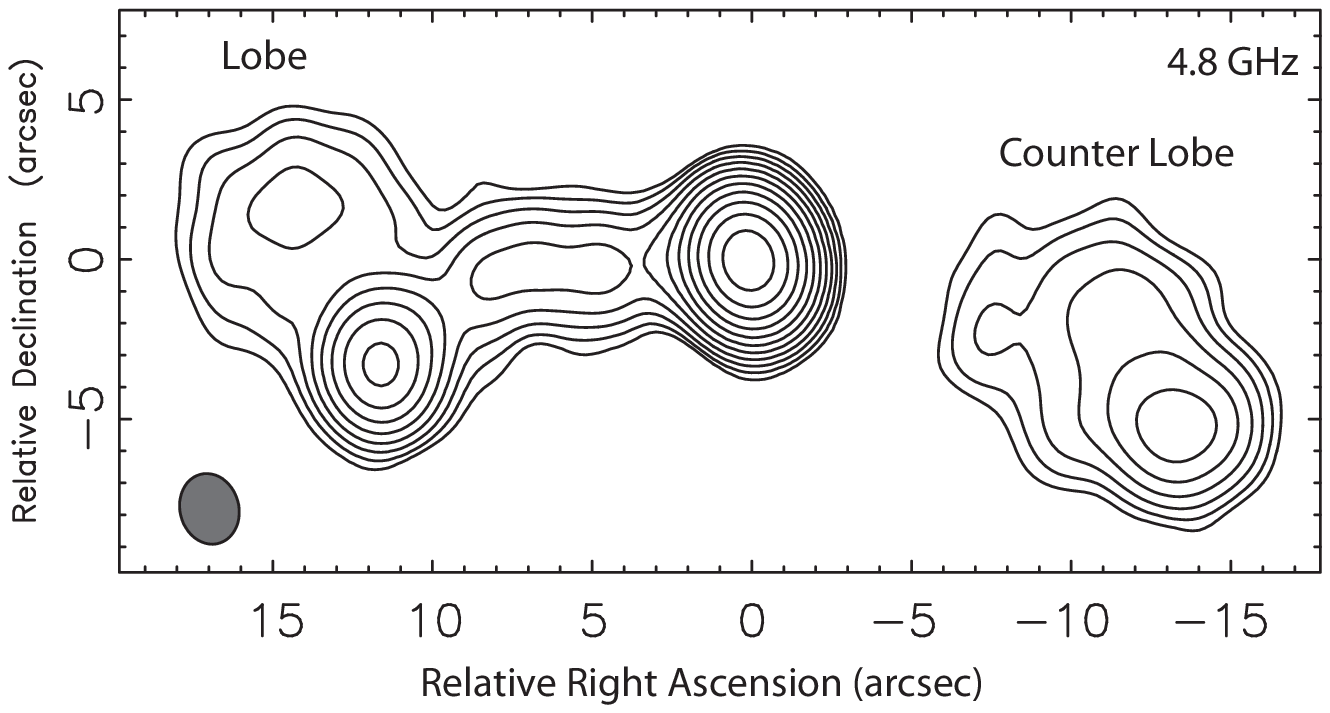}{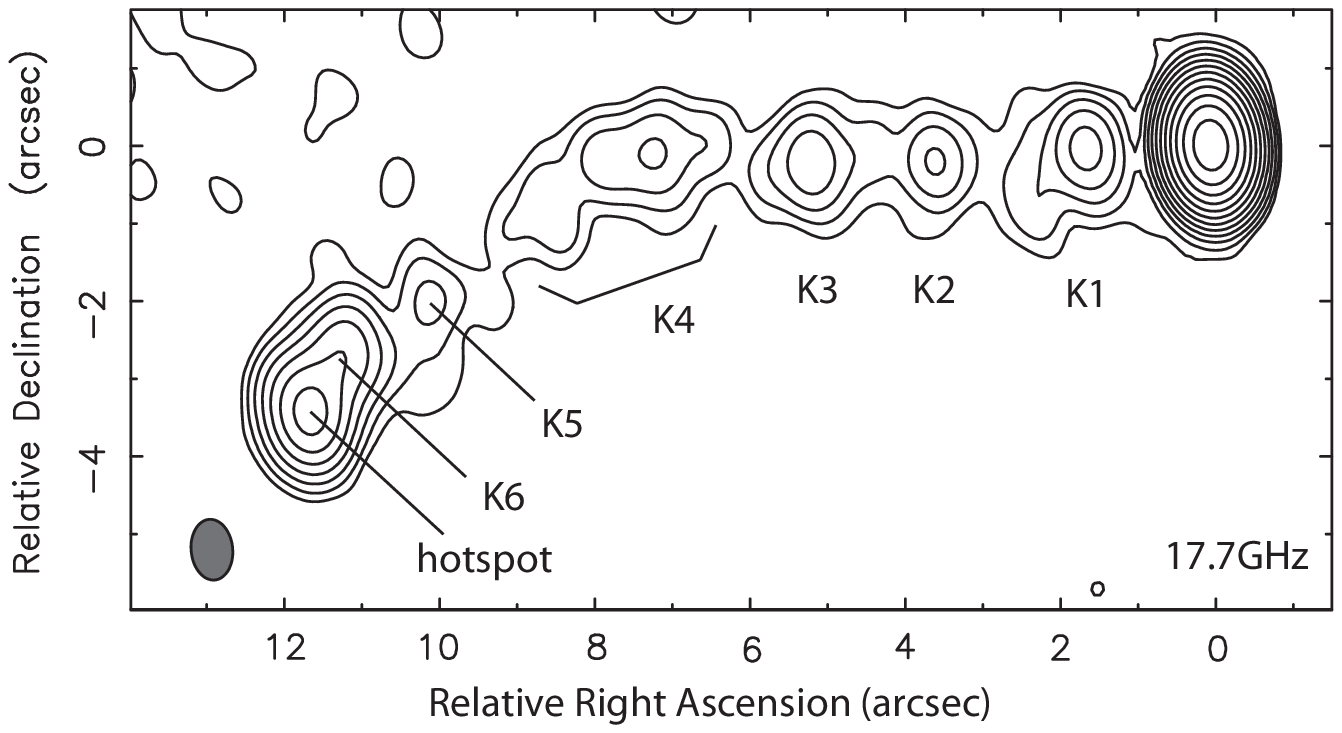}
\caption{ATCA images at 4.8~GHz (left) and 17.7~GHz (right) showing the radio structure of the jet and lobes of PKS~2101--490. Also shown is the naming convention used for various components of the jet. The scale of this image is $8.1$~kpc/$\arcsec$. Contours are separated by a factor of 2 in surface brightness. In the 4.8~GHz image, the lowest contour is 0.42 mJy/beam and the beam FWHM is $2\farcs24 \times 1\farcs84$. In the 17.7~GHz image, the lowest contour is 0.15 mJy/beam and the beam FWHM is $0\farcs79 \times 0\farcs54$. \label{fig:2101_radio_structure}}
\end{figure*}

\begin{figure*}[!ht]
\epsscale{0.8}
\plotone{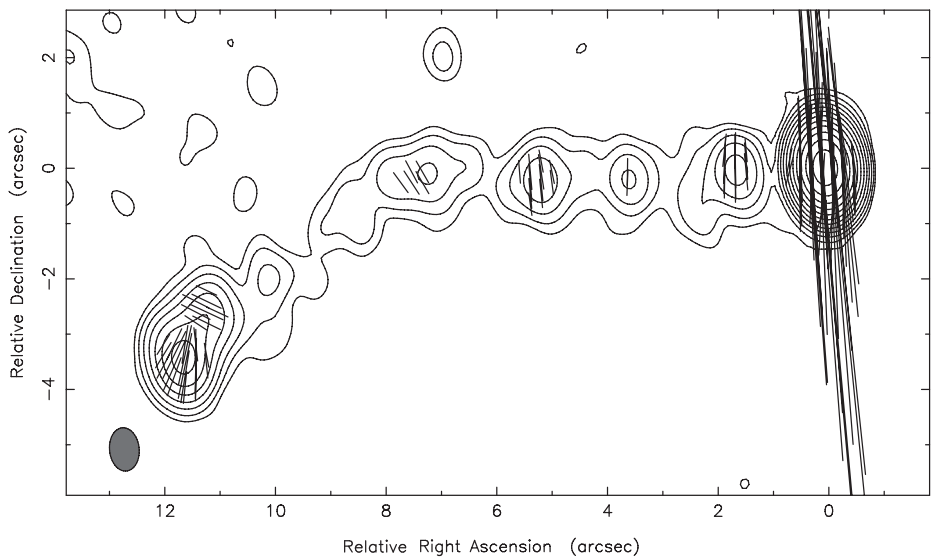}
\caption{Same as Figure \ref{fig:2101_radio_structure} (right) with polarization E-vectors overlaid. The E-vectors are scaled such that a length of 1 arcsecond corresponds to 1 mJy/beam polarized surface brightness. \label{fig:polarization_17GHz}} 
\end{figure*}

Figure \ref{fig:2101_radio_structure} illustrates the radio structure of the source and the naming convention used for the various features in the radio maps. Figure \ref{fig:polarization_17GHz} illustrates the polarization characteristics of the jet. We have extracted radio, optical and X-ray flux densities from the seven major emission regions identified in Figure \ref{fig:2101_radio_structure}. In this section we describe the observations in each waveband, as well as the methods used to measure flux densities and sizes for the individual jet knots. Table \ref{table:2101_obs_info} lists the observational information for all data used in this study.

\subsection{Radio}

PKS~2101--490 was observed with the ATCA at 4.8~GHz and 8.64~GHz in two configurations, 1.5A and 6A, on May 25 2000 and September 2 2000 respectively, and in a single configuration (6C) at 17.7~GHz and 20.2~GHz on May 10 2004. In each case, a full 12 hour synthesis was obtained, recording 128MHz bandwidth in all four polarization products. Regular scans on the nearby phase calibrator PKS~2106--413 were scheduled throughout the observations, as well as scans on the ATCA flux calibrator PKS~1934-638. Standard calibration and editing procedures were carried out using the MIRIAD data analysis package. Following calibration, the data were exported to DIFMAP and several imaging/self-calibration iterations were performed. The data were both phase and amplitude self-calibrated.

\begin{deluxetable*}{llllllllllll}
\tabletypesize{\tiny}
\tablecaption{Characteristics of Spatially Resolved Jet Knots and Lobes \label{table:knot_characteristics}}
\tablewidth{0pt}
\tablehead{
& \multicolumn{7}{c}{Flux Densities} &  & \multicolumn{3}{c}{De-convolved Dimensions} \\
\cline{2-8}  \cline{10-12} \\
Knot ID & F$_{4.8~\rm{GHz}}$ & F$_{8.6~\rm{GHz}}$ & F$_{17.7~\rm{GHz}}$ & F$_{20.2~\rm{GHz}}$ & F$_{806\rm{nm}}$ & F$_{475\rm{nm}}$ & F$_{1~\rm{keV}}$ & & $\phi_{\rm{Maj}\tablenotemark{a}}$ & $\phi_{\rm{Min}\tablenotemark{a}}$ & Vol\tablenotemark{b} \\
& mJy & mJy & mJy & mJy & nJy & nJy & nJy & & mas & mas & cm$^3$ \\
}
\startdata
Core & $720 \pm 40$ & $830 \pm 40$ & $500 \pm 50$ & $470 \pm 50$ & $2.32 \times 10^5$ & $1.59 \times 10^5$ & $87 \pm 3$ && --- & --- & --- \\
Knot 1 & --- & --- & $5.1 \pm 0.5$ & $4.7 \pm 0.5$ & $60 \pm 10$ & $30 \pm 10$ & $< 0.2$  && $400 \pm 100$ & $250 \pm 10$ & $3 \times 10^{65}$ \\
Knot 2 & --- & $4.6  \pm 0.5$ & $2.3 \pm 0.2$ & $2.2 \pm 0.2$ &$<$ 30 &$<$ 30 & $< 0.15$  &  & $400 \pm 100$ & $<$150  & $1 \times 10^{65}$ \\
Knot 3 & --- & $5.1  \pm 0.5$ & $3.5 \pm 0.3$ & $2.8 \pm 0.3$  &$<$ 30 & $<$ 30 & $0.5 \pm 0.15$ &  & $550 \pm 100$ & $330 \pm 100$ & $7 \times 10^{65}$ \\
Knot 4 & --- & $9 \pm 1$ & $4.7 \pm 0.5$ & $4.2 \pm 0.4$ &$<$ 50 & $<$ 50 & $ 1.3 \pm 0.3 $ & & $2500 \pm 500$ & $300 \pm 50$ & $3 \times 10^{66}$ \\
Knot 5 & --- & --- & $1.0 \pm 0.2$ & $0.8 \pm 0.2$ & $<30$ & $<30$ & $0.2 \pm 0.1$  & & $400 \pm 100$ & $300 \pm 100$ & $4 \times 10^{65}$ \\
Knot 6 & --- & $11 \pm 1$ & $6.5 \pm 0.6$ & $6.3 \pm 0.6$ & $180 \pm 10$ & $90 \pm 10$ & $0.75 \pm 0.2$  & & $600 \pm 50$ & $400 \pm 20$ & $1 \times 10^{66}$\\
hotspot & --- & $30 \pm 3$ & $16 \pm 1.5$ & $14 \pm 1.5$ & $<$ 30 &  $<$ 30  &$< 0.16$ & & $340 \pm 30$ & $240 \pm 20$ & $2 \times 10^{65}$ \\
Entire Jet & $115 \pm 1$ & $71 \pm 1$ & $40 \pm 0.5$ & $36 \pm 0.5$ & ---  & --- & $3.3 \pm 0.4$ & & --- & --- & --- \\
Lobe & $16 \pm 2$ & $8 \pm 1$ & $4.5 \pm 0.4$ & $4 \pm 0.4$ & --- & --- & $0.5 \pm 0.2$  & & 8600 & 6500  & $3 \times 10^{69}$ \\
Counter-Lobe & $65 \pm 6$ & $33 \pm 3$ & $17 \pm 2$ & $14 \pm 1$ & --- & --- & $1.5 \pm 0.2$  & & 12000 & 9400 & $9 \times 10^{69}$  \\
\enddata
\tablenotetext{a}{These are the FWHM Gaussian knot sizes, based on Gaussian fits to the radio data (except in the case of the lobes). In the case of K6 and the hotspot, the Gaussian fit was performed in the (u,v)-plane (see \S \ref{sec:radio_knot_measurements}). For all other jet knots, a Gaussian model was fit to the data in the image plane. In the case of the lobe and counter-lobe, the quoted sizes are simply the size of the flux extraction regions --- no fitting was done for the lobes.}
\tablenotetext{b}{These are apparent (projected) volumes calculated assuming ellipsoidal geometry, i.e. V = ($\pi$/6) $D_{\rm eq}^2 ~ D_{\rm pol}$ where $D_{\rm eq}$ is the equatorial diameter and $D_{\rm pol}$ is the polar diameter of the ellipsoid. If the knots are associated with stationary features in the jet, the de-projected volumes will be greater by a factor of ($1/\sin \theta$) where $\theta$ is the angle to the line of sight.}
\end{deluxetable*}

\subsubsection{Radio Knot Flux Density and Size Meausrements} \label{sec:radio_knot_measurements}

The spectrum of the entire jet (excluding lobe emission) between 4.8~GHz and 20.2~GHz is well described by a power law with spectral index $\alpha = 0.81 \pm 0.01$ (flux density $S_\nu \propto \nu^{-\alpha}$). Each of the four flux density measurements are within $1\%$ of the best fit power law (see Table \ref{table:knot_characteristics}), giving us confidence in the flux density and spectral index measurements for the entire jet. 
However, inspection of individual knot spectra indicates that the systematic uncertainties in flux density measurements for individual knots  are significantly greater than the off-source RMS. There are a number of factors contributing to the systematic uncertainty in the measurement of the relative strengths of individual knots, including the uniqueness problem that results from gaps in the (u, v)-coverage due to the small number of telescopes \citep[see e.g.][]{walker95}, and weak inter-knot emission that is detected at different levels in each band, as a result of the different (u, v)-coverage at each frequency. Also, some of the knots at 8.6~GHz are only marginally resolved, so that flux density measurements of individual knots becomes difficult. These issues, and estimation of systematic uncertainty are discussed in detail in \citet{godfreyPhD}. We are therefore unable to determine accurately the spectral index for each individual knot, and are instead forced to assume that each knot has the same spectral index as that of the entire jet, which has been determined accurately.

Knot sizes for jet components K1 - K4 were measured by fitting 2D Gaussian models to jet knots in the 17.7 and 20.2 GHz maps, using the AIPS task JMFIT. Knot 6 and the hotspot are partially blended even at 20~GHz, and for this reason we determined their flux density and size at 17.7 and 20.2 GHz by fitting elliptical Gaussian models to these components in the (u,v)-plane, using the \textit{modelfit} function of DIFMAP. The knot size estimates for all knots are based only on the measurements from the highest resolution maps (17.7 and 20.2 GHz). At 8.6 GHz, for knots 2, 3 and 4, flux densities were obtained by integrating the surface brightness within a region encompassing the knot emission. For knot 6 and the hotspot, the flux densities at 8.6 GHz were determined by fitting elliptical Gaussian models to these components in the (u,v)-plane. Due to the limited resolution at 8.6 Ghz, we were unable to accurately determine the flux density for knots 1 and 5.

The size estimates from the 17.7~GHz and 20.2~GHz images were consistent for some knots. However, in some cases the knot dimensions differed by as much as a factor of 2. This is another indication that image fidelity is questionable in the high frequency maps, and systematic errors resulting from the deconvolution/self-calibration process are significant. The knot sizes listed in Table \ref{table:knot_characteristics} are averages of the parameters determined from the 17.7~GHz and 20.2~GHz images. The uncertainties in knot size are taken to be half the difference between the 17.7 and 20.2~GHz measurements.

\subsection{Hubble Space Telescope Optical Observations}

PKS~2101--490 was observed with the Advanced Camera for Surveys (ACS) Wide Field Camera 1 (WFC1) on the Hubble Space Telescope in two filters (F475W and F814W) on March 8 2005. A total exposure time of 2.3 kiloseconds was obtained in each filter. A sub-pixel dithering pattern, with CR-SPLIT images at each of 3 positions along a chip diagonal, was utilized to eliminate bad pixels and allow us to maximize the angular resolution, as the ACS/WFC does not fully sample the PSF at either 4750 or 8000 \AA.  An ORIENT was chosen such that the jet did not fall within 25 degrees of a diffraction spike. The data were retrieved from the Multi-mission Archive at Space Telescope (MAST) website\footnote{HST archive \url{http://archive.stsci.edu/}}, however, multiple peaks in the pipeline drizzled image of the quasar core suggested that a re-reduction of the data was required. In addition, the pipeline does not take into account sub-pixel dithering, which was performed to in order to recover PSF information from the undersampling of the HST PSF. The re-reduction involved running MULTIDRIZZLE with a smaller PIXFRAC and scale, so that the images could be sub-sampled to 0.0247 arcsecond pixels, and checking the alignment of the images with TWEAKSHIFTS. The best reference files indicated in the HST archive were used for the re-reduction. The position of the optical quasar core was aligned with the position of the radio core. This required shifting the optical data by approximately $0\farcs4$.  

We used TinyTIM simulations \citep{krist94, suchkov98, krist04} to obtain subsampled PSF simulations for both bands.  For those simulations we assumed an optical spectrum of the form $F_\nu \propto \nu^{-1}$, although experience has shown that the PSF shape is not heavily dependent on spectral slope. Rotating the PSF to a north-up position for use with the drizzled images required independently rotating the two axes as the WFC detector's axes are not completely orthogonal on the sky.  We normalized the PSF to the total number of counts in a 2$\arcsec$ circular aperture centred on the source. This takes advantage of the fact that charge ``bleed" on the ACS is linear and charge is conserved for a saturated source \citep{gilliland04}, and enables optimal matching of the PSF's outer portion to what is observed. Because the quasar core was saturated in the individual exposures, this inevitably led to negatives in the central pixels when PSF subtraction was done; however, given the small residuals in other places plus the smooth off-jet isophotes, we believe the result is reliable.  Subtraction of the PSF allowed us to look for optical jet components within 1-2$\arcsec$, and resulted in a detection of Knot 1 in both bands.

\begin{figure*}[ht!]
\center
\includegraphics[scale=0.3925]{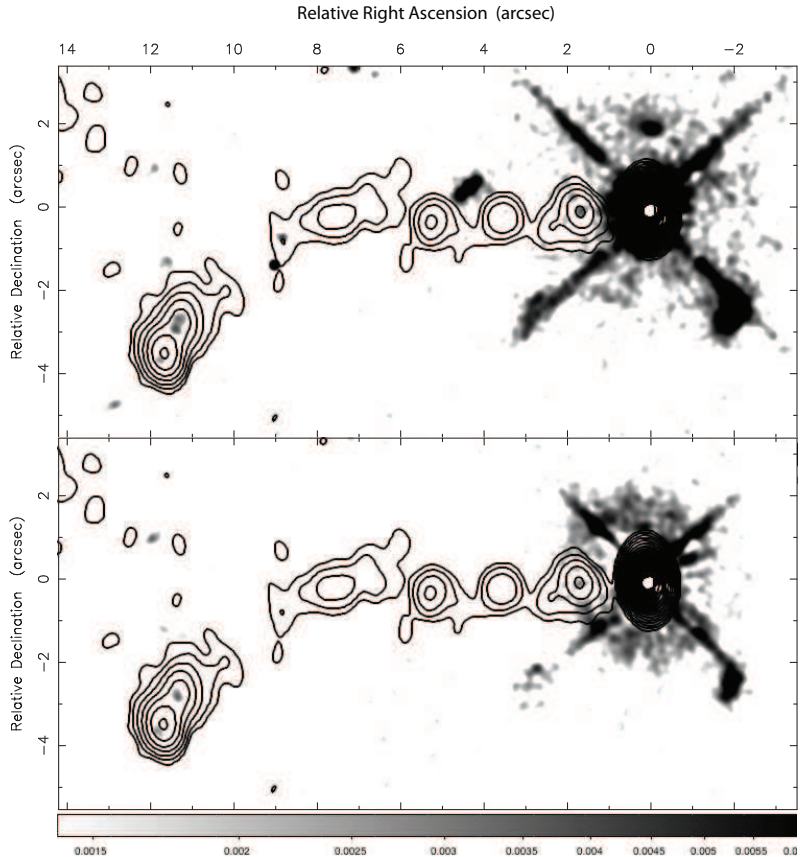}
\caption{Hubble Space Telescope images  of PKS~2101--490 (greyscales: Top = F814W; Bottom = F475W) with 20.2~GHz ATCA radio contours overlaid. The HST images have been sub-sampled with pixel size $0\farcs0247$, PSF subtracted from the quasar core, and smoothed with a $0\farcs3$ FWHM Gaussian to better show the optical counterpart to knot K6. The diffraction spikes are merely the residuals after PSF subtraction from the quasar core (see text). \label{fig:HST_radio_overlays}} 
\end{figure*}

Figure \ref{fig:HST_radio_overlays} shows the resulting optical maps with radio contours overlaid. In these images, the HST data have been smoothed with a 0.3 arcsecond FWHM Gaussian to better show the optical emission from knot K6, which is clearly detected in both filters. It is interesting to note that the F814W image of K6 reveals an elongated, double structure that is not apparent in the F475W image of K6. We further note that the position of the optical emission from K6 is coincident with the radio position to within the uncertainties associated with the optical-radio image alignment. We consider the interpretation of the optical data in \S \ref{sec:2101_K6_broken_powerlaw}.

\subsubsection{Optical Flux Density Measurements}

Knots 1 and 6 are the only jet features detected in the optical images. Optical flux densities were measured using standard aperture photometry techniques. The appropriate aperture corrections were taken from \citet{sirianni05} Table 3, and the appropriate extinction corrections were taken from \citet{sirianni05} Table 14 assuming E(B-V) = 0.039 at the position of PKS~2101--490\footnote{ This value for E(B-V) was obtained using the NASA Extragalactic Database extinction calculator \url{http://nedwww.ipac.caltech.edu/forms/calculator.html}, which is based on the Galactic reddening maps of \citet{schlegel98}.}. The optical flux densities and upper limits are given in Table \ref{table:knot_characteristics}.

\begin{deluxetable}{ccc}
\tabletypesize{\scriptsize}
\tablecaption{Knot 6 (K6) de-convolved sizes \label{table:K6_sizes}}
\tablewidth{0pt}
\tablehead{
\colhead{Band} & \colhead{Cross-jet\tablenotemark{a}} & \colhead{Along-jet\tablenotemark{a}} \\ \colhead{} & \colhead{(arcsec)} & \colhead{(arcsec)}
}
\startdata
ATCA 20.2~GHz & $0.4 \pm 0.02$ & $0.6 \pm 0.05$ \\
HST F814W & $0.24^{+0.03}_{-0.24}$  & $0.4 \pm 0.1$  \\
HST F475W & $0.24^{+0.03}_{-0.24}$ & $0.2^{+0.03}_{-0.2}$ \\
\emph{Chandra} 0.5 - 7~keV & $<$~0.6 &  $<$~0.6
\enddata
\tablenotetext{a}{These are the intrinsic FWHM Gaussian knot sizes, obtained via $\Theta_{\rm K6, intrinsic} = \sqrt{\Theta_{\rm K6}^2 - \Theta_{\rm PSF}^2}$, except in the case of F814W along-jet, which exhibits a double peak structure. In that case, the extent of the emission region was estimated by inspection of the longitudinal knot profile (Figure \ref{fig:optical_profile}).}
\end{deluxetable}

\subsubsection{The extent of optical emission from Knot 6}

\begin{figure}[ht!]
\center
\includegraphics[scale=0.5]{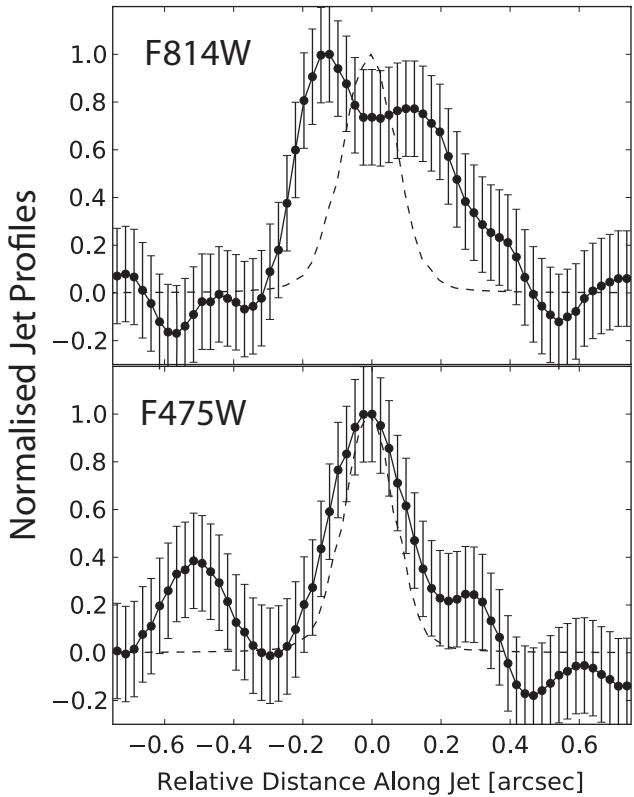}
\caption{Longitudinal jet profiles of knot 6 in both HST bands along a position angle of 150$^{\circ}$, integrated across a region of width $0\farcs5$. These plots serve to illustrate the difference in the apparent knot length in each band, and the relative alignment of the peaks. Note that the peak in the F475W profile lies between the peaks in the F814W profile. The HST maps were smoothed by a Gaussian with $0\farcs15$ FWHM before producing these profiles. The profile of the PSF, smoothed by a Gaussian with $0\farcs15$ FWHM, is represented by the dashed curve in each plot. The error bars are based on the rms of profiles made in off-source regions.  \label{fig:optical_profile}} 
\end{figure}

We estimated the size of the optical emission from K6 as follows: first, the optical images were convolved with $0\farcs15$ FWHM Gaussian. We then produced integrated profiles both parallel and perpendicular to the jet axis, for the knot K6, a nearby star, and the TinyTim generated PSF using the ds9 projection capability. From the integrated profiles, we measured the FWHM of each\footnote{The longitudinal profile in the F814W band appears double peaked, so in that case, we estimated the extent of the optical emission region simply via inspection of the profile (Figure \ref{fig:optical_profile}).} feature to be $\Theta_{\rm K6}$, $\Theta_{\rm star}$ and $\Theta_{\rm PSF}$. The profiles of the star and TinyTim PSF are in good agreement with each other, and indicate that in the smoothed maps, the PSF FWHM is $\Theta_{\rm PSF} \approx \Theta_{\rm star} \approx 0\farcs18$ for both the F475W and F814W bands. We then calculated the de-convolved (intrinsic) extent of the optical emission associated with the knot K6 as $\Theta_{\rm K6, intrinsic} = \sqrt{\Theta_{\rm K6}^2 - \Theta_{\rm PSF/Star}^2}$. The measured PSF sizes in the smoothed HST maps correspond to an un-smoothed PSF of $0\farcs11$ in both bands, as expected for ACS/WFC. 
We note that the profile of K6 along the jet in the F814W band appears double peaked, in contrast to the F475W profile which is single peaked (see Figures \ref{fig:HST_radio_overlays} and \ref{fig:optical_profile}).

The optical knot appears smaller than the associated radio knot: in both bands the cross-jet width of the optical emission associated with K6 is significantly less than the measured width in the radio band (Table \ref{table:K6_sizes}). The length of K6 parallel to the jet in the F814W map appears marginally resolved, while the length of K6 in the F814W map is clearly resolved, and appears to consist of two peaks (Figure \ref{fig:optical_profile}). A discussion of the optical emission from K6 is presented in \S \ref{sec:2101_K6_broken_powerlaw}.

\subsection{\emph{Chandra} X-ray Observations}

PKS~2101--490 was observed with the \emph{Chandra} X-ray observatory on 17 December 2004 (Cycle 6) using the Advanced CCD Imaging Spectrometer (ACIS-S) for a total exposure time of 42 kiloseconds (\dataset [ADS/Sa.CXO#obs/05731]
{Chandra ObsID 5731}). To reduce the effect of pile-up in the quasar core, a 1/4 subarray mode was used with a single CCD, so that the frame time was 0.8 seconds. The source was positioned close to the readout edge of the CCD to reduce the effect of charge transfer inefficiency. A new Event 2 data file was made with pixel randomization removed, and the X-ray and radio core positions were aligned, requiring a shift of the X-ray data by approximately $0\farcs4$. The data were restricted to the energy range 0.5 - 7~keV.

\begin{figure*}
\epsscale{1.1}
\plotone{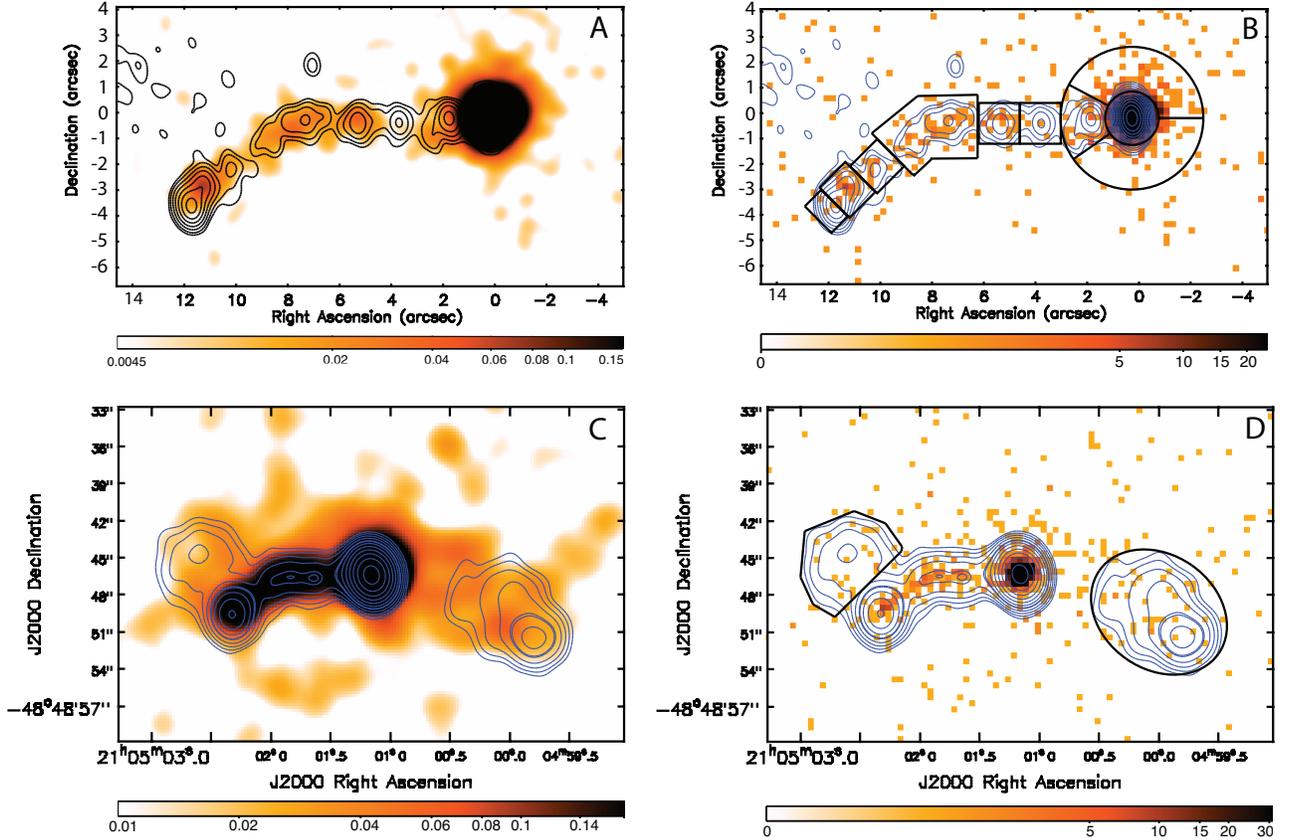}
\caption{(A) X-ray image binned to $0\farcs0492$ pixels and smoothed with a $0\farcs7$ FWHM Gaussian with 17.7~GHz ATCA contours overlaid. Colour scale is logarithmic between 0.0045 and 0.15 counts/beam. (B) Raw X-ray counts image binned to half a \emph{Chandra} pixel width ($0\farcs246$) with 17.7~GHz radio contours overlaid. Colour scale is logarithmic between 0 and 25 counts. The black outlines mark the flux extraction regions used for X-ray flux density calculations. (C) X-ray image binned to $0\farcs246$ pixels and smoothed with $0\farcs7$ FWHM Gaussian. Colour scale is logarithmic between 0.01 and 0.18 counts/beam. Blue contours are from the 4.8~GHz ATCA image. (D) Raw X-ray counts image with 4.8~GHz radio contours overlaid. Colour scale is logarithmic between 0 and 30 counts. The black outlines mark the flux extraction regions used to calculate X-ray flux densities associated with the lobes. \label{fig:X-ray_radio_overlays}} 
\end{figure*}

\subsubsection{X-ray Flux Density Measurements}

The X-ray flux densities for individual knots were obtained by calculating the background-subtracted counts in each knot region and multiplying by the conversion factor. The conversion factor was estimated by fitting an absorbed power law spectrum to the X-ray counts extracted from the whole jet. The X-ray spectra of the jet and lobe were fit using the Sherpa software package by minimising the Cash statistic (related to the log of the likelihood). The instrument response functions (RMF and ARF) were determined from the CALDB calibration database appropriate for the position of the source on the ACIS-S3 chip. In all model fits, the neutral Hydrogen column density was fixed at the Galactic value $3.4 \times 10^{20}$~cm$^{-2}$ as determined from the COLDEN\footnote{The COLDEN column density calculator, available at \url{http://cxc.harvard.edu/toolkit/colden.jsp}, based on \citet{dickey90}.} column density calculator provided by the \emph{Chandra} X-ray Center. The uncertainties for the flux density and spectral index were calculated using the ``Covariance" routine in Sherpa. The results of this routine are valid provided that the surface of Log Likelihood is approximately shaped like a paraboloid. The ``Interval-Projection" routine in Sherpa was used to verify that this condition was met. 

We extracted a total of 138 counts from the jet having energies in the range 0.5~keV to 7.0~keV. Fitting a power law gives 1~keV flux density $S = 3.3 \pm 0.4$~nJy and spectral index $\alpha_X = 0.85 \pm 0.2$. This implies the conversion factor is $G \approx 1.0 \> \mu$Jy/count/s for a spectral index of $\alpha \approx 0.85$. In the counter lobe we extracted a total of 55 counts in the energy range 0.5~keV to 7.0~keV. Following the same procedure as for the jet, we find flux density $S = 1.5 \pm 0.2$~nJy at 1~keV and spectral index $\alpha = 1.3 \pm 0.3$, implying the conversion factor is $G = 1.15 \> \mu$Jy/count/s for a spectral index $\alpha \approx 1.3$. These estimates of the conversion factor are consistent with the predictions of the \emph{Chandra} proposal planning toolkit.

Figure \ref{fig:X-ray_radio_overlays}A is a comparison of X-ray and radio structure in the jet of PKS~2101--490 and Figure \ref{fig:X-ray_radio_overlays}B illustrates the regions used to extract X-ray counts for the knots. The regions shown in Figure \ref{fig:X-ray_radio_overlays}B (except for the hotspot extraction region --- this region is discussed further below) have sides $\gtrsim 1\farcs6$. Knots 5, 6 and the hotspot are difficult to separate in the \emph{Chandra} image. In order to estimate the counts associated with the hotspot and avoid contamination from knot 6, a region encompassing only one side of the hotspot is used (the side furthest from knot 6). Due to the background in the vicinity of the hotspot and the possibility of contamination from knot 6, the few counts within this aperture may not be associated with the hotspot, and therefore the flux density estimate for the hotspot is an upper limit. An aperture correction of 2 is used when calculating the upper limit, since the flux extraction region encompasses only one side of the hotspot.

In the \emph{Chandra} image, knot 1 is blended with the wings of the X-ray core. To place a limit on the X-ray flux density from knot 1, a sector of an annulus centred on the core was used, as shown in Figure \ref{fig:X-ray_radio_overlays}B. The background was estimated using the section of the annulus excluding knot 1. The data are consistent with zero counts from knot 1.

\subsubsection{The extent of X-ray emission from Knot 6} \label{sec:K6_X-ray_knot_size}

The X-ray knot size is of great significance to models of jet X-ray emission \citep{tavecchio03}. Only knot 6 has sufficient counts and is sufficiently isolated from other strong knots to allow an accurate estimate of the knot size. Using the CIAO task \textit{dmlist}, events were extracted from a $0\farcs8$ radius circular aperture centered on knot 6. This region was chosen to be large enough to include a large fraction of the counts from knot 6, but small enough to avoid contamination from neighbouring regions of the jet. Note that the encircled energy fraction within a circular aperture of radius $0\farcs8$ is $\gtrsim 85 \%$ for a spectral index of $\alpha \sim 0.8$. A total of 29 events were extracted from this region. We assume, for simplicity, that the knot surface brightness profile and the inner $0\farcs8$ of the \emph{Chandra} PSF are both approximately Gaussian with variance $\sigma^2_{\rm K6}$ and $\sigma^2_{\rm PSF, \, 0\farcs8}$ respectively. We estimated the variance ($\sigma^2_{\rm PSF, \, 0\farcs8}$) of the \emph{Chandra} PSF within an $0\farcs8$ radius aperture by extracting events within an $0\farcs8$ circular aperture centered on the core, and calculating the standard deviation of event coordinates. Doing so, we find $\sigma_{\rm PSF, \, 0\farcs8} = 0\farcs29 \pm 0\farcs10$. This comparison between core and jet PSF is valid since the core and jet X-ray spectral indices are similar, and the core is not significantly affected by pile-up. We find that knot 6 is unresolved in both the jet and cross-jet directions.

To obtain an upper limit to the size of knot 6 we use the standard deviation distribution \citep{kennedy51}. We find that the 99$\%$ upper limit to the X-ray knot size is $\sigma_{\rm obs, \, upper} = 0\farcs4$. The upper limit to the de-convolved (intrinsic) size of knot 6 (taken as the Gaussian FWHM) is then $D_{K6}  < 2 \sqrt{2 \mbox{ln} 2} \left( \sigma^2_{\rm obs, \, upper} - \sigma^2_{\rm PSF, \, 0\farcs8} \right)^{1/2} = 0\farcs6$ in both the jet and cross-jet directions. Thus, the extent of the X-ray knot emission is less than or equal to the extent of the radio emission of knot 6.

\begin{figure*}[!ht]
\epsscale{0.65}
\plotone{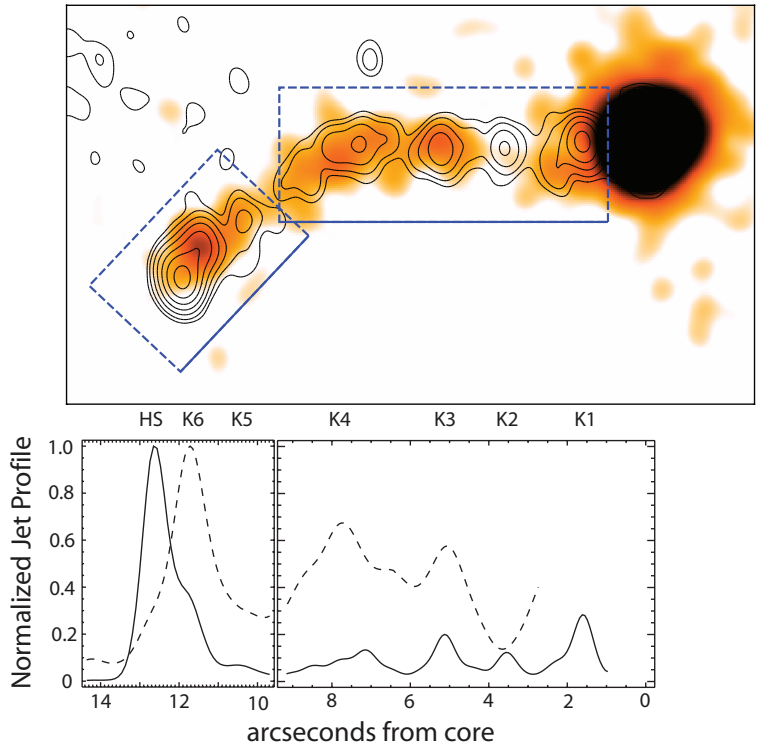}
\caption{Top: X-ray image binned with 1/10 \emph{Chandra} pixel width ($0\farcs0492$) and smoothed with a $0\farcs7$ FWHM Gaussian to emphasize the X-ray jet structure. 17.7~GHz ATCA contours are overlaid. Also shown are the projection regions used to obtain the radio and X-ray jet profiles. Bottom: X-ray (dashed line) and radio (solid line) longitudinal jet profiles. These curves show the jet brightness integrated across the jet as a function of distance from the core along the jet ridge-line. Note that in this plot, the X-ray resolution ($1\farcs1$) and radio resolution ($0\farcs65$) are not equal. These profiles simply serve to illustrate the differences in jet structure at each wavelength. \label{fig:radio_xray_profile}} 
\end{figure*}

\section{Results} \label{sec:results}

In this section we present a comparison of the radio and X-ray images; we describe the methods used to model the radio to X-ray SEDs of individual knots; and we present the results of spectral modeling. 

\subsection{Radio/X-ray longitudinal jet profiles}

Figure \ref{fig:radio_xray_profile} illustrates the radio and X-ray longitudinal jet profiles. The X-ray surface brightness is greatest near the end of the jet. This is in contrast to the trend that is seen in sources such as 3C273 \citep[e.g.][]{marshall01, sambruna01}, 0827+243 \citep{jorstad04} and 1127-145 \citep{siemiginowska02, siemiginowska07}, where the X-ray brightness peaks closer to the core and the brightness of the knots decreases with distance from the core \citep[see ][for a discussion of this phenomenon]{georganopoulos04}.

\subsection{Modeling the Spectral Energy Distributions of \\ Spatially Separated Knots}

The spectral energy distributions of the jet knots are typical of quasar X-ray jet knots such as those in PKS~0637--752: A single or broken power law is unable to fit the entire radio to X-ray SED, so that two spectral components are required to model the data (see Figure \ref{fig:SEDs}). The X-ray spectral index of the entire jet $\alpha_{0.5 \> \rm keV}^{7.0 \> \rm keV} = 0.85 \pm 0.2$ is consistent with the radio spectral index of the entire jet $\alpha_R = 0.81 \pm 0.01$. The data are therefore consistent with an inverse Compton interpretation for the X-ray emission.

\subsubsection{Synchrotron Self Compton Model} 

As with other quasar X-ray jets \citep[e.g.][]{schwartz00}, synchrotron self Compton (SSC) models for the X-ray jet emission in PKS2101-490 require sub-equipartition magnetic field strengths. The magnetic field strengths in the knots derived from SSC modeling are a factor of $\sim$50 below the equipartition (minimum energy) values \citep{godfreyPhD}. Such sub-equipartition magnetic fields are generally disfavoured, because in that case, the internal energy, internal pressure and jet power (for a fixed jet speed) are orders of magnitude greater than in the case of equipartition magnetic fields. We note that if the jet is Doppler beamed, then the ratio of the equipartition magnetic field strength to the SSC derived magnetic field strength increases approximately in proportion to $\delta^{2/(1+\alpha)}$.

\begin{deluxetable*}{ccccccc}
\tablecaption{IC/CMB Modeling Results \label{table:IC_modeling_results}}
\tabletypesize{\scriptsize}
\tablewidth{0pt}
\tablehead{
\colhead{Knot ID} & \colhead{$\frac{\rm{F}_{\rm 1 keV}}{\rm{F}_{\rm 17.7 \> GHz}}$} & \colhead{$\delta$} & \colhead{$B_{\rm eq}$} &  \colhead{$n_e$} & \colhead{$L^{e^{-}e^{+}}_{\rm jet}$} & \colhead{$L^{e^{-}p}_{\rm jet}$} \vspace{0.2cm} \\
& [$\times 10^{-8}$]  & & [$\mu G$] & [$\times 10^{-6}$ cm$^{-3}$] & [$\times 10^{46}$ ergs/s] & [$\times 10^{46}$ ergs/s] \vspace{0.2cm}
}
\startdata
Knot 1  	&$<4$			& $<6$ 		& $>40$ 		& $>0.6$		& $<0.5$		& $<3$ \\
&&&&&&\\

Knot 2  	&$< 6.5$ 			& $<7$ 		& $>40$	 	& $>0.5$	 	& $<0.2$ 		&$<1$  \\
&&&&&&\\

Knot 3  	&$14 \pm 4$ 		& $7$ 		& $30$ 		& $0.2$	 	& $0.5$ 		& $3$   \\
&&&&&&\\

Knot 4  	&$28 \pm 7$ 		& $7$ 		& $20$	  	& $0.1$		& $0.2$ 		& $1$   \\
&&&&&&\\

Knot 5  	&$20 \pm 10$ 		& $7$		& $20$ 		& $0.2$ 		& $0.3$		& $1$ \\
&&&&&&\\

Knot 6  	&$11 \pm 3$ 		& $6$		& $30$		& $0.3$	 	& $0.7$		& $4$   \\

\enddata
\tablecomments{Assumed model parameters: $\gamma_{\rm min} = 50$, $\gamma_{\rm max} = 10^5$, $\alpha = 0.8$, \\ Jet viewing angle = 9 degrees (for de-projection of knot volumes given in Table \ref{table:knot_characteristics}), \\ratio of proton to electron energy densities $\frac{\epsilon_p}{\epsilon_e} << 1$. \\ To calculate jet powers we have assumed $\Gamma = \delta$.}
\end{deluxetable*}

\subsubsection{IC/CMB Modeling}
\label{s:IC-CMB}

In this section we model the knot X-ray emission in terms of inverse Compton scattering of CMB photons in a highly relativistic jet directed close to the line of sight \citep{tavecchio00, celotti01}.

\paragraph{Assumed form of the electron energy distribution:} In order to apply the analytic one-zone IC/CMB model \citep{dermer95, harris02, worrall09}, we assume a single power law electron energy distribution $N(\gamma) = K_e \gamma^{-a}$ between a minimum and maximum Lorentz factor, $\gamma_{\rm min}$ and $\gamma_{\rm max}$, with $a = 2.6$ (corresponding to synchrotron spectral index $\alpha = 0.8$). We assume $\gamma_{\rm max} = 10^5$ for all knots. The assumed value for $\gamma_{\rm max}$ allows the synchrotron spectrum to cut-off at a frequency below that of the HST observing frequencies.  The value of the high energy cut-off is not well constrained, however, the conclusions drawn from this model are not sensitive to the assumed value of $\gamma_{\rm max}$ \citep[see][Appendix A]{schwartz06a}. In contrast to the single power law electron energy distribution assumed here, in \S \ref{sec:2101_K6_broken_powerlaw} we model the synchrotron spectrum of Knot 6 in terms of a broken power law. We note that the results of the IC/CMB model are insensitive to the details of the high energy end of the electron energy distribution, provided the break in the distribution occurs at $\gamma_{\rm break} >> \gamma_{\rm min}$. Therefore, the assumption of a single power law distribution in this section does not affect the conclusions drawn from this model for Knot 6, nor does it alter the conclusions drawn in later sections.

The low energy cutoff, $\gamma_{\rm min}$, is constrained so that the model does not produce optical IC/CMB emission above the HST upper limits or detections. This constraint on $\gamma_{\rm min}$ is possible because an extrapolation of the IC/CMB spectrum from X-ray to optical frequencies lies above the optical upper limits, and in the case of Knot 6, is comparable to the detection level but with an incorrect spectral index. In addition to this constraint, $\gamma_{\rm min}$ must not be so high that there is no significant X-ray emission produced at 0.5~keV. As we show in the following sections, the IC/CMB model requires a Doppler factor $\delta \sim 6$ at least in some parts of the jet. Assuming $\nu_{\rm min}^{\rm ic/cmb} \approx 1.6 \times 10^{11} \times \delta^2 \gamma_{\rm min}^2$ Hz \citep[see e.g.][]{worrall09}, we are able to constrain the value for $\gamma_{\rm min}$ to lie in the approximate range $10 \lesssim \gamma_{\rm min} \lesssim 200$. \citet{mueller09} find marginal evidence for a value $\gamma_{\rm min} \approx 50$ in the jet of PKS~0637--752 based on spectral fitting the jet X-ray spectrum. We therefore adopt $\gamma_{\rm min} = 50$ in the following analysis. 

\paragraph{The model and assumptions:} 

We use the standard one-zone IC/CMB model equations \citep[see][]{worrall09} assuming a continuous jet geometry (i.e. $S_\nu \propto \delta^{2 + \alpha}$ for the synchrotron flux density). Many of the jet knots appear elongated along the jet axis. Therefore, since this model requires the jet viewing angle to be small, we take account of projection effects in the calculation of knot volumes. Without independent constraints on the jet viewing angle, we simply assume a representative value in order to make an approximate de-projection. As we show in the following sections, the IC/CMB model requires a Doppler factor of order $\delta \gtrsim 6$ at least in some parts of the jet. This implies that the jet viewing angle must be $\lesssim 9$ degrees. Angles significantly less than 9 degrees are unfavourable since that would imply uncomfortably large source size. For example, if the jet lies closer than 5 degrees to the line of sight, the total source size must be $> 2.3$ Mpc. Therefore, we assume a jet viewing angle of $\theta = 9$ degrees in making an approximate de-projection of the knot length along the jet axis, and note that this angle corresponds approximately to the maximum jet viewing angle given the derived Doppler factor $\delta \gtrsim 6$. The de-projected volumes, V, are related to the projected volumes $V_0$ as $V = V_0 / \sin \theta$. The projected volumes $V_0$ are listed in Table \ref{table:knot_characteristics}. For the purposes of the current calculations, and to reduce the number of model parameters, we further assume that the proton contribution to the particle energy density is negligible. 

The results of IC/CMB modeling are presented in Table \ref{table:IC_modeling_results}. Included in the table is the jet kinetic energy flux, which we calculate based on the derived jet parameters in the case of purely leptonic and electron/proton jets, using equation B17 of Appendix B in \citet{schwartz06a}, with the following simplifying assumptions: The bulk Lorentz factor of the jet $\Gamma >> 1$, equipartition between particle and magnetic field energy densities, and a tangled magnetic field geometry such that $\langle B^{\prime 2}_{\perp} \rangle = (2/3) B^2$. With these assumptions, the expression for kinetic luminosity in the case of a purely leptonic and electron/proton jet respectively, are (in c.g.s. units)
\begin{eqnarray}
L_{jet}^{e^{+/-}} &\approx& \pi R^2 \Gamma^2 c \left(  \frac{B^2}{3 \pi}  \right) \quad \rm{ergs/s} \nonumber \\
L_{jet}^{e^{-} p} &\approx& \pi R^2 \Gamma^2 c \left( \left(  \frac{\Gamma - 1}{\Gamma} \right) n_e m_p c^2 +  \frac{B^2}{3 \pi}  \right) \quad \rm{ergs/s} \nonumber
\end{eqnarray}
In the case of an electron/proton jet, we assume one cold proton for each relativistic electron. The electron density $n_e$, in c.g.s. units, is calculated from the spectral fits as 
\begin{equation}
n_e \approx \frac{B^2}{8 \pi m_e c^2} \left(  \frac{a-2}{a-1}  \right)  \gamma_{\rm min}^{-1}
\end{equation}
where the magnetic field strength B is in Gauss. 

The model-dependent results presented in Table \ref{table:IC_modeling_results} suggest that, if the results of this model are correct, while the jet Lorentz factor remains approximately constant along the jet, the magnetic field and particle density decrease by a factor of a few between the innermost knots and the outer knots. The results indicate that there is modest, if any, loss of kinetic luminosity along the jet.

\subsection{Comparison of jet morphology with PKS~0637--752 and PKS~0920--397}

The jet of PKS~2101--490 undergoes an apparent bend of approximately 45$^\circ$ at about half way between the core and jet termination, somewhat reminiscent of the jet morphology in PKS~0637--752 \citep{schwartz00}. However, unlike PKS~0637--752, the X-ray emission in PKS~2101--490 is detected after the jet bend. This is of interest because a change in jet viewing angle produces a change in Doppler factor, which should manifest itself as change in knot brightness and the X-ray to radio flux density ratio. However, there is not a unique relationship between apparent bend angle and change in jet viewing angle. Therefore, the fact that X-ray emission continues beyond the bend in 2101-490 cannot be used to constrain the emission mechanism. It may simply be the case that the bend in 0637--752 is associated with a significant increase in viewing angle (and a hence significant decrease in the Doppler factor) whilst the bend in 2101--490 may be associated with a relatively small change, or a decrease in viewing angle (and hence small change, or increase in Doppler factor). Having said that, on a population basis, the probability distribution of change in jet viewing angle for a given apparent bend angle may enable a useful constraint on the emission mechanism. 


In addition to the morphological similarities to PKS~0637--752, we note a striking similarity between PKS~2101-490 and PKS~0920-397 \citep{schwartz10}, particularly in the vicinity of the jet termination. In both these sources, a bright, compact knot approximately 1 arcsecond ($>$ 10~kpc de-projected) upstream from the jet termination is detected in radio, optical and X-ray bands. A similar, X-ray bright pre-hotspot jet knot is seen in 1354+195 \citep{sambruna02}. One possible interpretation is that these pre-hotspot jet knots may be associated with an oblique shock that is produced as the jet enters the highly turbulent region at the head of the cocoon. Numerical simulations of extragalactic jets show that as the jet approaches the hotspot, it encounters an increasingly violent environment, with strong turbulence that can perturb the jet flow, inducing instabilities or directly causing oblique shocks to form due to density/pressure gradients in the lobes \citep[eg.][]{norman96}. The pre-hotspot jet knots may be the result of the jet entering an increasingly violent environment near the hotspot.

\section[Comparison of the Jet and Hotspot]{Comparing Independent Estimates of Jet Energy Flux for PKS~2101--490} \label{sec:jet_power}

In this section, as a means to assess the validity of the IC/CMB model, our goal is to obtain independent estimates of jet power and compare these with the estimate of jet power derived from the IC/CMB model.

\subsection{Jet Energy Flux from Hotspot Parameters}

It is possible to estimate the jet power based on observed hotspot properties by applying the conservation of momentum and making a number of reasonable assumptions about the hotspot. In the following section we carry out this approach, and compare the hotspot derived jet power estimates to those obtained from IC/CMB modelling of Knot 6 --- the jet feature closest to the hotspot.

Consider a uniform jet of area A, particle energy density $\epsilon$, pressure p, mass density $\rho$,  relativistic enthalpy $w = \epsilon + p + \rho c^2$, magnetic field perpendicular to the flow direction $B_{\perp}$ (as indicated by the polarization), speed $c \beta$ and corresponding bulk Lorentz factor $\Gamma$. The kinetic power ($L_{\rm jet}$) and momentum flux ($F_M$) along the jet are, in c.g.s. units \citep[e.g.][]{double04}:

\begin{eqnarray}
L_{\rm jet} &=& A \Gamma^2 \beta c \left( w + \frac{B_{\perp}^2}{4 \pi}  \right) \qquad  \label{eqn:F_E} \\
F_M &=& A \left[ \Gamma^2 \beta^2 \left(  w + \frac{B_{\perp}^2}{4 \pi}   \right)  + p + \frac{B_{\perp}^2}{8 \pi} \right] \qquad \label{eqn:F_M}
\end{eqnarray} 
Let us consider first the region of the jet upstream of the hotspot. In this region
the IC/CMB model indicates that the jet Lorentz factor, $\Gamma >> 1$.
In such a highly relativistic jet the kinetic power $L_{\rm E, jet}$ is simply related to the momentum flux $F_{\rm M, jet}$ via 
\begin{eqnarray}
L_{\rm jet} &\approx& c \times  F_{\rm M, jet}
\end{eqnarray}
We assume a near normal shock at the jet terminus and appeal to conservation of momentum, so that $F_{\rm M, jet} = F_{\rm M, hotspot}$, and thus $L_{\rm jet} \approx c~F_{\rm M, hotspot}$. Assuming that the jet plasma in the hotspot is decelerated to a low velocity so that $\Gamma^2 \beta^2 \ll 1$, the following relation then holds:
\begin{eqnarray}
L_{\rm jet} &\approx& c A \times \left[ p + \frac {B_\perp^2}{8 \pi} \right]_{\rm hs} \label{eqn:F_E_vs_F_M}
\end{eqnarray}
The above equality holds regardless of assumptions about the jet characteristics such as its composition or the ratio of magnetic to particle energy densities in the jet or hotspot. We estimate the jet kinetic luminosity simply by estimating the hotspot parameters. This technique of jet energy flux estimation will be considered further in an upcoming paper (Godfrey \& Shabala, in prep.).
 
In order to estimate hot spot parameters, we assume that the lepton population is ultra-relativistic ($p = \epsilon/3$) and dominates the particle pressure. Then
\begin{equation}
L_{\rm jet} \approx A c \frac{B^2}{8 \pi} \left( 1 + \frac{1}{3} \frac{\epsilon_{e^{\pm}}}{\epsilon_B} \right)
\end{equation}

As a check on this analysis, we apply it to the case of Cygnus A. Synchrotron self Compton modelling of Cygnus A hotspot A, assuming a power law electron energy distribution of the form $N(\gamma) = K_e \gamma^{-a}$, indicates B=150 $\mu$G and B$_{\rm eq}$ = 280 $\mu$G assuming R=2kpc and $a = 2.1$ \citep{wilson00}. For hotspot D: B=150 $\mu$G and B$_{\rm eq}$ = 250 $\mu$G assuming R=2.2kpc and $a = 2.05$. \citet{wilson00} do not not give an estimate of the electron energy density, so we estimate the electron energy density from the SSC and equipartition magnetic field strengths as $\epsilon_e = \frac{B_{\rm eq}^2}{8 \pi} \left(  \frac{B}{B_{\rm eq}} \right)^{\left(a+1 \right)/2}$, since $\epsilon_e \propto K_e$ and the synchrotron flux density $S_\nu \propto K_e B^{(a+1)/2} \Rightarrow \epsilon_e \propto B^{-(a+1)/2}$, in the case of a power law distribution of the form $N(\gamma) = K_e \gamma^{-a}$. Therefore $\epsilon_e = 8 \times 10^{-9}$~ergs/cm$^{3}$ and $\epsilon_e = 5.4 \times 10^{-9}$~ergs/cm$^{3}$ for hotspots A and D respectively. Hence, we find $L_{\rm jet} \approx 1 \times 10^{46}$~ergs/s for both hotspots A and D. This estimate of the jet power in Cygnus A is in excellent agreement with the value obtained by \citet{wilson06}, $L_{\rm jet} \gtrsim 1.2 \times 10^{46}$~ergs/s, based on modeling the cocoon dynamics. An independent estimate of jet power in Cygnus A comes from \citet{lobanov98} who show that frequency dependent shifts of the radio core enable a determination of the jet parameters, which, when applied to the case of Cygnus A, gives a value for the jet power $L_{\rm jet} = (0.55 \pm 0.05) \times 10^{46}$ ergs/s.

\begin{deluxetable}{ccc}
\tabletypesize{\scriptsize}
\tablecaption{Jet Power derived from IC/CMB modelling of K6 SED \label{table:jet_powers}}
\tablewidth{0pt}
\tablehead{
\colhead{Composition} & \colhead{$\gamma_{\rm min} = 10$} & \colhead{$\gamma_{\rm min} = 50$} 
}
\startdata
Leptonic & $1 \times 10^{46}$~ergs/s & $7 \times 10^{45}$~ergs/s \\
Electron/Proton & $3 \times 10^{47}$~ergs/s & $4 \times 10^{46}$~ergs/s \\ 
\enddata
\end{deluxetable}

We now return to the case of PKS~2101--490. The \emph{Chandra} X-ray image provides only an upper limit to the hotspot X-ray flux density. Therefore, to calculate the hotspot magnetic field strength and particle energy density, we adopt equipartition estimates. We note that the adoption of an equipartition magnetic field strength does not overproduce hotspot X-ray emission, for which we have obtained an upper limit. We justify the assumption of equipartition by the fact that high luminosity hotspots typically exhibit X-ray flux density consistent with the equipartition synchrotron self Compton model predictions \citep{hardcastle04}. Moreover, the total energy density, and hence derived jet power, is only weakly dependent on magnetic field strength for magnetic fields near equipartition conditions, so that moderate departure from equipartition does not alter our conclusions. For example, if the hotspot magnetic field strength were 1/3 the equipartition value, then our equipartition jet power would under-estimate the true value by less than a factor of 2.

The hotspot synchrotron spectral index is not well constrained, so we apply a single power law model along with the following assumptions: $10 < \gamma_{\rm min} < 1000$; $\gamma_{\rm max} > 10^4$; $\alpha = 0.8$. This model, combined with the flux densities and volume listed in Table \ref{table:knot_characteristics} indicates that $B_{\rm eq, hs} \sim 200 - 450 \> {\rm \mu G}$. Due to the prior assumption of equipartition we have $\epsilon_{e^\pm} = \epsilon_{\rm B}$.
Taking $A = (6 \pm 1) \times 10^{43}$~cm$^{2}$ as estimated from the 20~GHz ATCA image, the jet energy flux calculated from the hotspot model is
\begin{equation}
L_{\rm jet} = (0.4 - 2) \times 10^{46} \> \rm{ergs/s}
\end{equation}
The range of values reflects the range of $\gamma_{\rm min}$ assumed in the hotspot model, with the lower jet power corresponding to higher assumed values of $\gamma_{\rm min}$. Minimum Lorentz factors $\gamma_{\rm min} \approx 600$ have been observed in a number of hotspots \citep[][and references therein]{godfrey09}. This is an order of magnitude greater than the value of $\gamma_{\rm min}$ assumed in the jet. As suggested by \citet{godfrey09}, such an increase in $\gamma_{\rm min}$ between the jet and hotspot may be due to the dissipation of jet energy in a relativistic proton/electron jet terminating at a near-normal shock.

\subsection{Willott et al. $Q_{\rm jet} - L_{\rm 151}$ relation}

A widely used method for estimating jet power in high luminosity radio sources is based on the model of \citet{willott99}, in which the jet power is estimated using the 151 MHz radio luminosity as follows:
\begin{equation}
Q_W \approx f^{3/2} 1.5 \times10^{38} \left(  \frac{S_{151} D_L^2}{10^{28}~\rm{W~HZ^{-1}~sr}^{-1}} \right)^{6/7} \> {\rm W}
\end{equation}
where $Q_W$ is the kinetic power per jet, $S_{151}$ is the flux density of the entire source at 151 MHz, $D_L$ is the luminosity distance, and f is a parameter that accounts for uncertainties in various model assumptions, with $1 < f \lesssim 20$ \citep[see][for details of the model]{willott99}. For high power sources, the value of f has not been constrained empirically, but is typically taken to be $f = 10 - 20$ \citep[e.g.][]{fernandes11, hardcastle07}. We extrapolate flux density measurements of $S_{\rm 408~MHz} = 2.34$~Jy \citep{large81}, $S_{\rm 843~MHz} = 1.48$~Jy \citep{mauch03} and $S_{\rm 1.41~GHz} = 1.1$~Jy \citep{wright90} to estimate the flux density at 151 MHz --- $S_{\rm 151~MHz} \approx 4.25$~Jy --- and thereby obtain $Q_W \sim 1~(3) \times 10^{46}$~ergs/s assuming $f=10~(20)$. 

\subsection{Average jet power from lobe parameters and source age}

Here we estimate the average jet power over the life of the source based on the lobe parameters and an estimate of the source age.
\begin{equation}
L_{\rm jet} \approx \frac{2~U_{\rm lobe}}{t_{\rm age}}
\end{equation}
where $U_{\rm lobe}$ is the total (magnetic plus particle) energy contained in the lobe. The factor of 2 is an approximate correction accounting for the fraction of jet energy converted to lobe kinetic energy and work done by the expanding lobes \citep[][]{willott99, bicknell97}. In \S \ref{sec:lobes} we estimate the lobe parameters via synchrotron and inverse Compton modelling of the radio to X-ray SED, and argue that the source age is likely $t_{\rm age} \sim 10^8$~yrs. Based on the results of \S \ref{sec:lobes} we estimate the average jet power over the life of the source to be 
$L_{\rm jet} \gtrsim 10^{45} \> \mbox{ergs/s.}$
This is a factor of 4 - 20 lower than the other estimates of jet power. We note that we have ignored the possible contribution of protons to the energy density of the lobes, and we have ignored adiabatic cooling in estimating the cooling rate in the above analysis, each of which would increase the jet power estimate. We note that the calculations in \S \ref{sec:lobes} involve projected volumes. Using de-projected volumes will not increase the total total energy, since the product $K_e \times V$ remains constant, and the particle energy dominates $U_{\rm lobe}$ in this source.

\begin{figure*}[!ht]
\epsscale{0.8}
\plotone{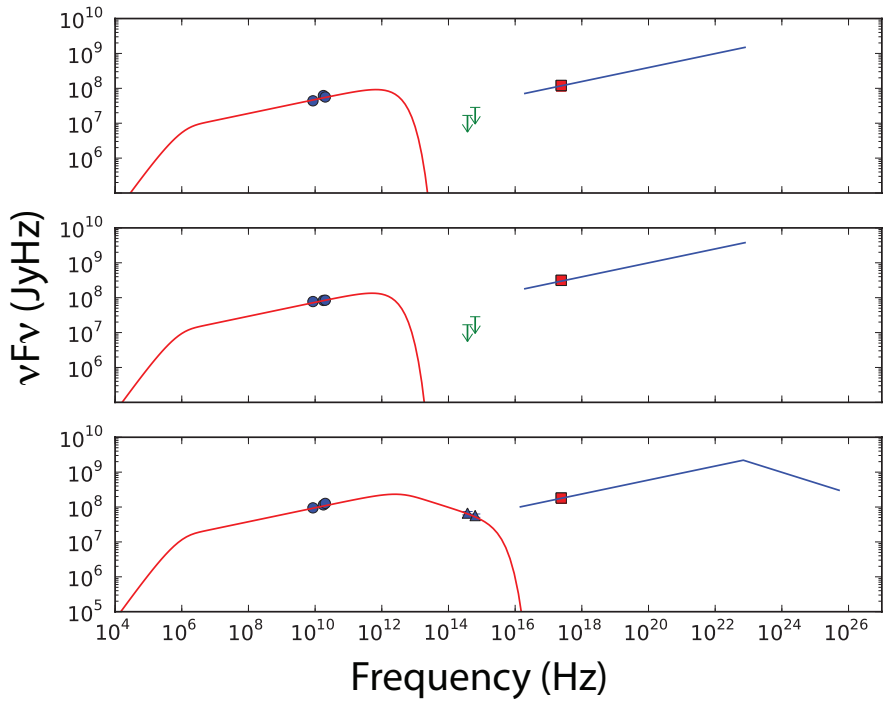} 
\caption{Spectral energy distributions for knots 3 (top), 4 (middle) and 6 (bottom). Also plotted are the synchrotron and analytic power law model for the IC/CMB interpretation, with $\alpha = 0.8$, $\gamma_{\rm min} = 50$, and $\gamma_{\rm max} = 10^5$, and in the case of knot 6, a broken power law model is used (see \S \ref{sec:2101_K6_broken_powerlaw} for details). The parameters for the fits are given in Table \ref{table:IC_modeling_results}. Note that while the X-ray spectral index of individual knots could not be measured, the X-ray spectral index of the entire jet is $\alpha_{\rm 0.5 \> keV}^{\rm 7 \> keV} = 0.85 \pm 0.2$, consistent with the slope of the model in the X-ray band. \label{fig:SEDs}} 
\end{figure*}

\begin{figure}[!ht]
\epsscale{1.0}
\plotone{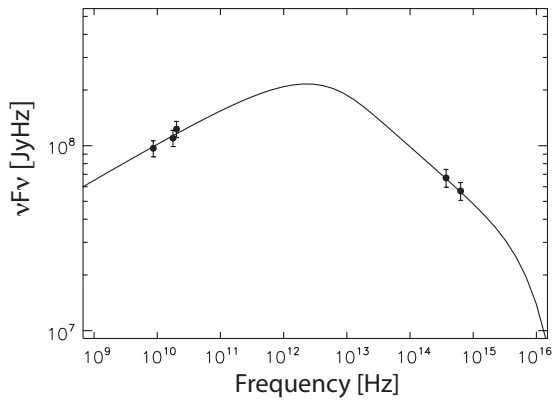}
\caption{Broken power law synchrotron spectrum fit to the radio to optical spectral energy distribution of knot 6. The relevant parameters of the fit are: $\alpha = 0.8$, $\nu_b = 7 \times 10^{12}$~Hz, $\nu_{\rm max} = 10^{16}$~Hz. }
\label{fig:K6_broken_power_law_fit}
\end{figure}

\subsection{Discussion of the Results}

Three independent order-of-magnitude estimates of jet power indicate $L_{\rm jet} \sim 10^{45} - 10^{46}$~ergs/s. For comparison, the jet power obtained from the one-zone IC/CMB model of jet X-ray emission are given in Table \ref{table:jet_powers}, for different values of the minimum Lorentz factor $\gamma_{\rm min}$, and different assumed jet compositions.

It can be seen that, in the case of a purely leptonic composition, there is good agreement between the IC/CMB model and independent estimates of jet power, regardless of the value for $\gamma_{\rm min}$ within the stated limits. Under the assumption of a cold proton/electron composition, there is good agreement between the IC/CMB model and independent estimates of jet power, provided $\gamma_{\rm min} \gtrsim 50$. This assumes one cold proton per radiating electron. If relativistic protons are included, the lower limit on $\gamma_{\rm min}$ would increase. 

We have shown that in the case of PKS~2101--490, the IC/CMB predicted jet power is $<< 10^{48}$~ergs/s (the ``uncomfortably large" jet power derived for some quasar X-ray jets based on the IC/CMB model \citep{atoyan04, mehta09}) under a range of reasonable assumptions. Indeed, the IC/CMB derived jet power is comparable to independent estimates of $L_j \sim 10^{45} - 10^{46}$~ergs/s for both leptonic and electron/cold proton compositions, provided $\gamma_{\rm min} \gtrsim 50$. 
An important point to note about our jet power estimates is that, for the IC/CMB model, we have used a cylindrical geometry for knot K6 with length approximately 10 times its width. This highly elongated cylindrical geometry is required by the observation that knot K6 is elongated along the jet axis in the radio maps, and the fact that the jet must be aligned within $\lesssim 9$ degrees of the line of sight if the IC/CMB model is to be applied. Typically, jet knots are assumed to be spherical with the knot diameters equal to the jet widths \citep[e.g.][]{tavecchio00, mehta09}. If we were to assume a spherical volume in our analysis, the IC/CMB derived jet powers would be greater by a factor of $\sim$ 4.


\section{The Optical Emission from Knot 6}  \label{sec:2101_K6_broken_powerlaw}

\subsection{Introduction}

The rate of optical jet discovery\footnote{See http://astro.fit.edu/jets/ for a listing of known optical/IR jet sources, and information for each source.} in objects with confirmed radio and X-ray jets is approximately 2 out of 3. However, few of these sources have resolved optical knots, particularly at high redshift \citep[excepting PKS~0637--752 in which multiple optical knots are resolved:][]{schwartz00, mehta09}, and whilst a systematic study has not yet been performed, it would appear that relatively few have detailed information regarding the spectral slopes in both the radio and optical bands.

In PKS~2101--490, knot 6 is the only jet region that is reliably detected in all three bands (radio, optical and X-ray). The radio through optical spectrum for knot 6 can be fit using a broken power law with a break of $\Delta \alpha = 0.5$ (Figure \ref{fig:SEDs}). This suggests that radiative cooling may be responsible for the steeper spectrum at optical frequencies. However, the interpretation of the optical emission from Knot 6 is complicated by fact that the optical to X-ray spectral index  is $\alpha_{6.32 \times 10^{14} \rm Hz}^{\rm 1~ keV} \approx 0.8$, consistent with the observed radio and X-ray spectral index of the entire jet. It is therefore plausible that a significant fraction of the observed optical flux density is IC/CMB emission. For the IC/CMB model to produce the observed optical flux at $6.32 \times 10^{14}$ Hz, the electron energy distribution must continue with the same slope to Lorentz factors $\gamma \lesssim 10$, assuming $\Gamma \sim \delta = 6$. If the electron distribution cuts off at a Lorentz factor $\gamma >> 10$, the IC/CMB emission will not make a significant contribution to the observed optical flux density. 

With this caveat in mind, in this section we model the radio through optical spectrum in terms of a broken power law synchrotron model, and consider whether the jet parameters derived from IC/CMB modeling are consistent with this interpretation of the radio to optical spectrum of knot 6. Figure \ref{fig:K6_broken_power_law_fit} illustrates the broken power law fit to the SED of knot 6. The parameters of this fit are $\alpha = 0.8$, $\nu_{\rm b} = 7 \times 10^{12}$~Hz and $\nu_{\rm max} = 10^{16}$~Hz. 

\subsection{Synchrotron model for the radio to optical spectrum of Knot 6} \label{sec:K6_synchrotron_model}

In modeling the radio to optical spectrum we consider the effects of synchrotron cooling \citep{meisenheimer89} and adopt the following procedure: We assume that relativistic electrons are injected at a shock with an energy index $a$ (number density per unit Lorentz factor, $N(\gamma) = K_e \gamma^{-a}$). The electrons cool as a result of synchrotron and inverse Compton emission downstream of the shock over a length $D$, which defines the extent of the emitting region. In this model we assume that the post-shock magnetic field and velocity are uniform; this is a reasonable assumption for an oblique shock. An electron with Lorentz factor $\gamma$ cools as a result of synchrotron and inverse Compton emission according to:
\begin{equation}
\frac {d\gamma}{dt^\prime} = - \xi \, \gamma^2
\end{equation}
where $t^\prime$ is time in the plasma co-moving frame, 
\begin{equation} \label{eqn:A}
\xi = \frac {4}{3} \frac {\sigma_T}{m_e c} \left( \frac {B^2}{8 \pi} + U_{\rm CMB} \right) \, ,
\end{equation}
in c.g.s. units, and the energy density of the cosmic microwave background in the plasma co-moving frame is given by 
\begin{equation}
U_{\rm CMB} = U_{CMB,0} \, (1+z)^4 \, \Gamma^2
\end{equation}
where the current epoch CMB energy density is $U_{\rm CMB,0} \approx 4.2 \times 10^{-13} \> \rm ergs \> cm^{-3}$ and $\Gamma$ is the jet Lorentz factor. 

The integrated emission from the post-shock region is the superposition of the emission from a number of progressively cooled slices. The volume averaged number density per unit Lorentz factor is described by:
\begin{equation} \label{eqn:N_gamma}
\bar{N}(\gamma)  = \left \{ \begin{array}{ll} 0 &  \gamma < \gamma_{\rm{min}}
  , \quad \gamma > \gamma_{\rm{max}} \\ 
 \frac{K_e \gamma_b}{(a-1)} \gamma^{-(a+1)} g\left( \frac{\gamma}{\gamma_b}
 \right) & \gamma_{\rm{min}}  < \gamma < \gamma_{\rm{max}} \\ 
\end{array}
\right.
\end{equation}
where
\begin{equation} \label{eqn:g_gamma}
g \left( \frac{\gamma}{\gamma_b} \right)  = \left \{ \begin{array}{ll}  1 -
  \left(1 - \frac{\gamma}{\gamma_b} \right)^{a-1}    &  \gamma < \gamma_b \\ 
1 &  \gamma > \gamma_b \\
\end{array}
\right.
\end{equation}
The break Lorentz factor $\gamma_b$ is given by:
\begin{equation}
\gamma_b = \frac {c}{\xi D} \Gamma_{\rm sh} \beta_{\rm sh}
\label{e:gamma_break}
\end{equation}
where $\beta_{\rm sh}$ is the velocity of plasma relative to the shock and $\Gamma_{\rm sh}$ is the corresponding Lorentz factor. The electron Lorentz factor $\gamma_b$ is the Lorentz factor to which an electron of initially infinite energy cools following the shock. If the jet-shock is stationary then $\Gamma_{\rm sh} = \Gamma$.

The electron distribution (\ref{eqn:N_gamma}) describes a broken power law spectrum with the electron
spectral index smoothly changing from $a$ to $(a+1)$ at $\gamma \approx
\gamma_b$. The corresponding synchrotron spectrum is a broken power law with spectral index smoothly changing from $\alpha$ to $\alpha + 0.5$ at a frequency in the observer's frame, 
\begin{equation}
\nu_b = \frac{\delta}{(1+z)} \frac{3}{4 \pi} \Omega_0 \gamma_b^2
\label{e:nu_break}
\end{equation}
where $\Omega_0 ( = \frac{q_e B}{m_e c}$ in c.g.s. units) is the non-relativistic gyro-frequency. This equation relates the break frequency to the break Lorentz factor, and it is the break frequency, $\nu_b$, that is actually used in the fit; the other parameter is the injected spectral index, $\alpha = (a-1)/2$. The values for the magnetic field $B \approx 30 \> \mu \rm G$ and Doppler factor $\delta \approx 6$ have previously been determined from the relationship between X-ray and radio emission in \S~\ref{s:IC-CMB}. In order to minimize the number of parameters we assume that $\Gamma = \delta$ in the following analysis.

We can compare the fitted break frequency with the value theoretically implied by equations~(\ref{e:gamma_break}) for 
$\gamma_b$ and (\ref{e:nu_break}) for $\nu_b$. The theoretical break frequency is
\begin{equation}
\nu_b = \frac {27}{64 \pi} \, \frac {\delta}{1+z} \, \frac {\Omega_0 m_e^2 c^4}{D^2 \sigma_T^2} \, 
\frac{\Gamma_{\rm sh}^2 \beta_{\rm sh}^2}{U^2} 
\label{e:nu_b-pred}
\end{equation}
where $U=U_{\rm B} + U_{\rm CMB}$ is the total magnetic plus radiation energy density. The jet parameters derived from the IC/CMB model for this knot imply that inverse Compton losses dominate over synchrotron losses, since the CMB energy density in the jet rest frame is greater than the magnetic energy density. In order to compare the predicted break frequency with that determined by the synchrotron model we need to estimate the length, $D$, of the emitting region. The radio image of knot 6 is found to be elongated in the jet direction with de-convolved (intrinsic) length $D \approx 0\farcs6$. The upper limit to the size of the X-ray knot is $0\farcs6$. The maximum allowed viewing angle for $\delta = 6$ is $\theta_{\rm max} \approx 9^{\circ}$, therefore, we take the de-projected knot length to be of order $D \approx 0\farcs6/\sin \theta_{\rm max} \approx 4\arcsec \approx 30 \> \rm{kpc} \approx 9 \times 10^{22}$~cm. If we adopt $D \approx 10^{23} \>$cm and insert in equation~(\ref{e:nu_b-pred}) the appropriate parameter values from the IC/CMB fit  we obtain a predicted break frequency of $1.2 \times 10^{13} \> \rm Hz$. This is less than a factor of two above the modelled value of $7 \times 10^{12} \> \rm Hz$.
Hence, the jet parameters derived from IC/CMB modeling of knot 6 are broadly consistent with the break frequency estimated by fitting a broken power law to the radio and optical data points. 

We note that the F814W HST image of knot 6 (Figure \ref{fig:HST_radio_overlays}) contains a double-peaked appearance, suggesting that a one-zone model of the emission region, which spans $\gtrsim$30 kpc in the radio band, may not be applicable.

\begin{figure}[!ht]
\epsscale{1.15}
\plotone{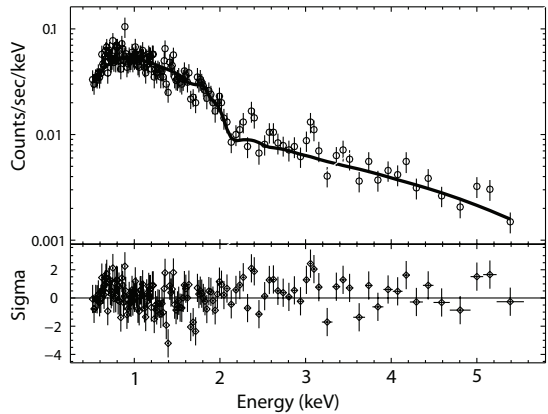}
\caption{Observed ACIS-S spectrum of the core of PKS~2101--490 with model spectrum (absorbed power-law) overlaid. Also shown are the residuals of the fit, expressed as [(data - model)/error] in each bin. The positive residuals at approximately 3.1 keV may be associated with a 6.4 keV Fe K$\alpha$ line redshifted by z=1.04, consistent with the spectroscopic redshift \citep[][Gelbord \& Marshall, in prep.]{marshall05}. \label{fig:core_X-ray_spectrum}} 
\end{figure}

\section{X-ray Spectral Analysis of the Quasar Core} \label{sec:core}

X-ray events were extracted using a circular aperture of radius 2$\arcsec$ centred on the core peak position. A total of 3687 events were extracted with energies in the range 0.5-7~keV. The pile-up fraction is estimated using the PIMMS Proposal Planning Toolkit to be $2\%$, and is therefore neglected in the following analysis. The pulse-height amplitude spectrum was re-binned to have a minimum of 20 counts per bin, so that the $\chi^2$ statistic could be applied. 

The spectrum was fit using an absorbed power law model, with HI column density fixed at the Galactic value $N_{\rm HI} = 3.41 \times 10^{20}$~cm$^{-2}$ as determined from the COLDEN column density calculator provided by the \emph{Chandra} X-ray Center. The data, along with the best fit curve are shown in Figure \ref{fig:core_X-ray_spectrum}. The parameters of the best fit are $\alpha_{\rm X-ray} = 0.84 \pm 0.05$,  $S = 87 \pm 3$~nJy at 1~keV, and  chi-squared per degree of freedom $\chi^2$ / d.o.f. = 146.4 / 147 $\approx$ 1.00. 

\citet{marshall05} reported a redshift of $z = 1.04$ for this source, based on an unpublished Magellan spectrum (see \S \ref{sec:intro}). We note positive residuals consistent with a 6.4 keV Fe K$\alpha$ line redshifted by $z\approx1.04$. If we include a line in the chi-squared fit, the model parameters become $\alpha_{\rm X-ray} = 0.85 \pm 0.05$,  $S = 87 \pm 3$~nJy at 1~keV, and the chi-squared per degree of freedom becomes $\chi^2$ / d.o.f. = 134.7 / 145 $\approx$ 0.93. Although an F-test would indicate that the line is significant, \citet{protassov02} show that such a test is not applicable to adding a spectral line to our basic model, and we have not carried out the extensive Monte Carlo testing they recommend since the jet physics is the primary result of this paper.

\section{Lobes} \label{sec:lobes}

Within the lobes, the radiation energy density of the CMB ($U_{\rm CMB} = 7 \times 10^{-12}$~ergs~cm$^{-3}$) is greater than the energy density of the locally generated synchrotron radiation. We therefore model the X-ray emission from the radio lobes in terms of inverse Compton scattering of the CMB, in this case assuming $\delta = 1$. In modeling the radio lobe emission we assume a broken power law electron energy distribution of the form given by equation \ref{eqn:N_gamma}. The value for the break Lorentz factor $\gamma_{b}$ is constrained by the X-ray spectrum. We argue as follows. The radio spectral index of the lobe is $\alpha_{\rm r} = 1.05 \pm 0.1$. We therefore assume that the radio emission corresponds to the post-break region of the electron spectrum. We assume a low-energy electron index, $a=2.1$, so that the slope of the model synchrotron spectrum above the break matches the observed radio spectral index. The X-ray spectral index in the lobe is $\alpha_{\rm X-ray} = 1.3 \pm 0.3$. The inverse Compton X-ray emission in the \emph{Chandra} X-ray band (0.5 - 7~keV) is produced by $\gamma \approx 600 - 2000$ electrons, given that the scattered photons are CMB photons with an average energy $\approx 3.6 \> \gamma^2 \> kT_{\rm CMB} \> (1+z)$, where $T_{\rm CMB} \approx 2.7 \>\rm K$ is the current epoch temperature of the cosmic microwave background. The steep X-ray spectrum therefore indicates that the Lorentz factor corresponding to the cooling break must be $\gamma_b \lesssim 600$. Hence, we adopt $\gamma_{\rm b} = 500$. Given the unknown jet viewing angle, we assume projected volumes in the analysis. If the jet is angled close to the line of sight, as required by the IC/CMB model for jet X-ray emission, the volumes may be underestimated by a factor of 5 - 10. Larger volumes would imply closer agreement with the equipartition model. 

\begin{figure*}[!ht]
\epsscale{0.99}
\plotone{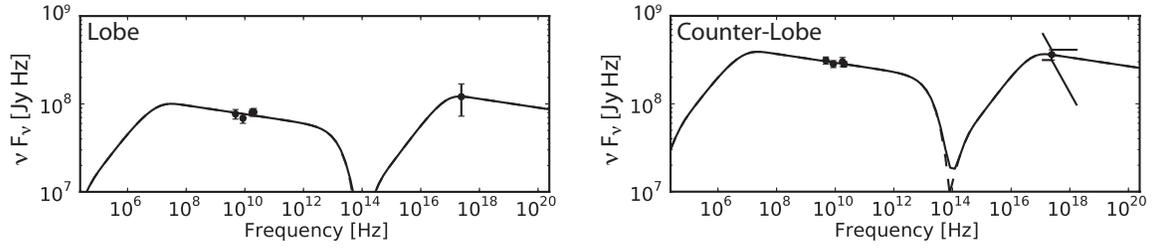}
\caption{Lobe and counter-lobe spectral energy distributions. The solid lines are the best fit synchrotron plus IC/CMB model. The parameters of the best fit models are given in table \ref{table:lobe_fit_results}. The ``bow-tie" around the counter-lobe X-ray data point illustrates the constraint on the X-ray spectral index of $\alpha_{\rm x} = 1.3 \pm 0.3$. There are not enough X-ray counts in the lobe to provide meaningful constraints on the spectral index. However, in modelling the lobe SED we have assumed the same X-ray spectral index as the counter-lobe. \label{fig:lobe_SED}} 
\end{figure*}

Figure \ref{fig:lobe_SED} shows the model fit to the radio and X-ray data for the lobe and counter lobe. The best-fit parameter values are given in Table \ref{table:lobe_fit_results}, along with the minimum energy magnetic field strength calculated using standard expressions \citep[e.g.][]{worrall06}. The magnitude of the average magnetic field strength, as well as the ratio of B to B$_{\rm min}$ is typical of powerful radio galaxies \citep{croston05}.

\begin{deluxetable*}{ccccccccc}
\tabletypesize{\scriptsize}
\tablecaption{Parameters of model fits to the lobe spectra. \label{table:lobe_fit_results}}
\tablewidth{0pt}
\tablehead{
& Volume & \colhead{$\alpha$} & \colhead{$\gamma_{\rm min}$} & \colhead{$\gamma_{\rm b}$} & $\gamma_{\rm max}$ & $K_e$ & B & $B_{\rm min}$  \\
& [cm$^3$] & & & & & [cm$^{-3}$] & [$\mu$G] & [$\mu$G] \\
}
\startdata
Counter-lobe & $9 \times 10^{69}$ & 0.55 & 10 & 500 & $6 \times 10^5$ & $1 \times 10^{-5}$ & 14 & 29 \\
Lobe & $4 \times 10^{69}$ & 0.55 & 10 & 500 & $6 \times 10^5$ & $0.8 \times 10^{-5}$ & 13 & 25 \\
\enddata
\end{deluxetable*}

The presence of a low value for the break Lorentz factor ($\gamma_b \lesssim 500$) indicates an old source. However, enhanced inverse Compton cooling leads to a shorter cooling time than for radio galaxies at the current epoch. The cooling time for electrons of Lorentz factor $\gamma$ in a magnetic field $B$ (and hence magnetic energy density $U_{\rm B}$) and immersed in the microwave background radiation with energy density $U_{\rm CMB}$ is 
\begin{eqnarray}
t_{\rm cool} &=& \frac {m_e c}{\sigma_T} \> U_{\rm CMB}^{-1} \left( 1 + \frac {U_{\rm B}}{U_{\rm CMB}} \right)^{-1} \, \gamma^{-1} 
\nonumber \\
&\approx& 3.6 \times 10^8 \> \left( 1 + \frac {U_{\rm B}}{U_{\rm CMB}} \right)^{-1} \> 
\left( \frac {\gamma}{500} \right)^{-1} \> \rm yrs
\end{eqnarray}
for a microwave background temperature $(1+z) \times 2.725 \> \rm K$. For the estimated lobe magnetic field strength, $U_{\rm B} \approx U_{\rm CMB}$, and $\gamma_b \sim 500$, implying a source age $t_{\rm age} \lesssim 2 \times 10^8 \> \rm yrs$. Adiabatic losses would increase the cooling rate, meaning the above prediction may be an over-estimate of the source age, hence the inequality in the estimate of $t_{\rm age}$ given above. For a deprojected jet length $\gtrsim 500$~ kpc (viewing angle $\lesssim 9^{\circ}$ as required by the beamed, equipartition IC/CMB model for jet X-ray emission) this implies a source expansion  speed $\gtrsim 0.01 \, c$. This estimate does not differ greatly from the estimate of the typical expansion speeds of lobes of a few percent of the speed of light \citep{scheuer95}.

\section{Conclusions} \label{sec:conclusions}

We have presented an analysis of new \emph{Chandra}, HST, and ATCA images for the quasar jet source PKS~2101--490. We extracted the radio to X-ray spectral energy distributions from seven regions of the 13$\arcsec$ jet, and modeled the jet X-ray emission in terms of Doppler beamed inverse Compton scattering of the cosmic microwave background (IC/CMB) for a jet in a state of equipartition between magnetic and particle energy densities (\S \ref{sec:results}). Using this model, we derived a bulk Lorentz factor $\Gamma \sim 6$ and magnetic field strength of order 30~$\mu$G. 

A major goal of this work has been to assess the validity of the beamed, equipartition IC/CMB model for jet X-ray emission through the use of additional constraints: namely, independent estimates of jet power, and the location of an apparent cooling break in the synchrotron spectrum of one of the jet knots. Below we discuss the results of our analysis for each of these additional constraints. 

The jet power predicted by the Doppler beamed, equipartition IC/CMB model was found to be in good agreement with independent order-of-magnitude estimates of jet power for this object, provided that $\gamma_{\rm min} \sim 50$ in the jet, and the knots are significantly longer than the jet width, as implied by de-projection of the observed knot lengths (\S \ref{sec:jet_power}). 

The brightest X-ray knot was detected in two HST filters, and the radio to optical data points were modelled as a broken power law with a standard cooling break $\Delta \alpha = 0.5$. The inferred break frequency was found to be consistent with the break frequency predicted using the IC/CMB model parameters along with the de-projected knot length that is implied by the small jet viewing angle required by the IC/CMB model (\S \ref{sec:2101_K6_broken_powerlaw}). However, we noted sub-structure in the F814W HST image of the bright optical knot, which is incompatible with a simple one-zone continuous injection model. 

Finally, we note positive residuals consistent with a 6.4 keV Fe K$\alpha$ line redshifted by the estimated z=1.04 reported by \citet{marshall05}. 

\acknowledgments

L.E.H.G would like to thank the Grote Reber Foundation for financial support during part of this work. H.L.M., E.S.P. and J.G. acknowledge the support of HST Grant GO-10352. DAS is supported by NASA contract NAS8-03060 and CXC grant GO9-0121B. The Australia Telescope Compact Array is part of the Australia Telescope which is funded by the Commonwealth of Australia for operation as a National Facility managed by CSIRO. We gratefully acknowledge the anonymous referee for their comments which helped to improve and clarify the manuscript.

{\it Facilities:} \facility{ATCA, Chandra, HST}


\vspace{2cm}

\begin{thebibliography}{}

\bibitem[Atoyan 
\& Dermer(2004)]{atoyan04} Atoyan, A., \& Dermer, C.~D.\ 2004, \apj, 613, 151 


\bibitem[Bicknell et al.(1997)]{bicknell97} Bicknell, G.~V., 
Dopita, M.~A., \& O'Dea, C.~P.~O.\ 1997, \apj, 485, 112 


\bibitem[Celotti et al.(2001)]{celotti01} Celotti, A., 
Ghisellini, G., \& Chiaberge, M.\ 2001, \mnras, 321, L1 


\bibitem[Chartas et al.(2000)]{chartas00} Chartas, G., et al.\ 
2000, \apj, 542, 655 

\bibitem[Croston et al.(2005)]{croston05} Croston, J.~H., 
Hardcastle, M.~J., Harris, D.~E., Belsole, E., Birkinshaw, M., 
\& Worrall, D.~M.\ 2005, \apj, 626, 733 



\bibitem[Dermer(1995)]{dermer95} Dermer, C.~D.\ 1995, \apjl, 
446, L63 

\bibitem[Dickey \& Lockman(1990)]{dickey90} Dickey, J.~M., \& Lockman, F.~J.\ 1990, \araa, 28, 215 

\bibitem[Double et al.(2004)]{double04} Double, G.~P., Baring, 
M.~G., Jones, F.~C., \& Ellison, D.~C.\ 2004, \apj, 600, 485 

\bibitem[Ekers(1969)]{ekers69} Ekers, J.~A.\ 1969, Australian 
Journal of Physics Astrophysical Supplement, 7, 3 

\bibitem[Fernandes et al.(2011)]{fernandes11} Fernandes, C.~A.~C., 
et al.\ 2011, \mnras, 411, 1909 



\bibitem[Gelbord et al.(2005)]{gelbord05} Gelbord, J.~M., Marshall, H.~L., Worrall, D.~M., et al.\ 2005, \apjl, 632, L75 

\bibitem[Georganopoulos 
\& Kazanas(2004)]{georganopoulos04} Georganopoulos, M., \& Kazanas, D.\ 2004, \apjl, 604, L81 

\bibitem[Georganopoulos et al.(2005)]{georganopoulos05} Georganopoulos, 
M., Kazanas, D., Perlman, E., \& Stecker, F.~W.\ 2005, \apj, 625, 656 



\bibitem[Gilliland(2004)]{gilliland04} Gilliland, R.~L.\ 2004, Instrument Science Report ACS 2004-01, 18 pages, 17 

\bibitem[Godfrey et al.(2009)]{godfrey09} Godfrey, L.~E.~H., et 
al.\ 2009, \apj, 695, 707 

\bibitem[Godfrey(2009)]{godfreyPhD} Godfrey, L.~E.~H. \ 2009, 
Ph.D.~Thesis, Australian National University

\bibitem[Hardcastle et al.(2004)]{hardcastle04} Hardcastle, M.~J., 
Harris, D.~E., Worrall, D.~M., \& Birkinshaw, M.\ 2004, \apj, 612, 729 

\bibitem[Hardcastle(2006)]{hardcastle06} Hardcastle, M.~J.\ 2006, 
\mnras, 366, 1465 

\bibitem[Hardcastle et al.(2007)]{hardcastle07} Hardcastle, M.~J., 
Evans, D.~A., \& Croston, J.~H.\ 2007, \mnras, 376, 1849 

\bibitem[Harris 
\& Krawczynski(2002)]{harris02} Harris, D.~E., \& Krawczynski, H.\ 2002, \apj, 565, 244 

\bibitem[Harris 
\& Krawczynski(2006)]{harris06} Harris, D.~E., \& Krawczynski, H.\ 2006, \araa, 44, 463 

\bibitem[Jester et al.(2006)]{jester06} Jester, S., Harris, 
D.~E., Marshall, H.~L., \& Meisenheimer, K.\ 2006, \apj, 648, 900 

\bibitem[Jorstad 
\& Marscher(2004)]{jorstad04} Jorstad, S.~G., \& Marscher, A.~P.\ 2004, \apj, 614, 615 

\bibitem[Kataoka 
\& Stawarz(2005)]{kataoka05} Kataoka, J., \& Stawarz, {\L}.\ 2005, \apj, 622, 797 

\bibitem[Kennedy \& Keeping(1951)]{kennedy51} Kennedy, J.~F., \& Keeping, E.~S.\ 1951, in The Mathematics of Statistics, Part 2, ed. N.J. Van Norstrand  

\bibitem[Krist \& Burrows(1994)]{krist94} Krist, J. E., \& Burrows, C. J., 1994, WFPC2 ISR 94-01

\bibitem[Krist \& Hook(2004)]{krist04} Krist, J. E., \& Hook, R., 2004, ``The TinyTim UserÕs Guide, Version
6.3", http://www.stsci.edu/software/tinytim/tinytim.pdf

\bibitem[Large et al.(1981)]{large81} Large, M.~I., Mills, 
B.~Y., Little, A.~G., Crawford, D.~F., 
\& Sutton, J.~M.\ 1981, \mnras, 194, 693 

\bibitem[Lobanov(1998)]{lobanov98} Lobanov, A.~P.\ 1998, \aap, 330, 79 

\bibitem[Lovell(1997)]{lovell97} Lovell, J.~E.~J.\ 1997, 
Ph.D. thesis, Univ. Tasmania

\bibitem[Lovell et al.(2000)]{lovell00} Lovell, J.~E.~J., Tingay, S.~J., Piner, B.~G., et al.\ 2000, Astrophysical Phenomena Revealed by Space VLBI, 215 

\bibitem[Marshall et al.(2001)]{marshall01} Marshall, H.~L., et 
al.\ 2001, \apjl, 549, L167

\bibitem[Marshall et al.(2005)]{marshall05} Marshall, H.~L., et 
al.\ 2005, \apjs, 156, 13 

\bibitem[Marshall et al.(2011)]{marshall11} Marshall, H.~L., et 
al.\ 2011, \apjs, 193, 15 


\bibitem[Massaro et al.(2009)]{massaro09} Massaro, F., Harris, 
D.~E., Chiaberge, M., et al.\ 2009, \apj, 696, 980 



\bibitem[Massaro et al.(2011)]{massaro11} Massaro, F., Harris, 
D.~E., \& Cheung, C.~C.\ 2011, \apjs, 197, 24 


\bibitem[Mauch et al.(2003)]{mauch03} Mauch, T., Murphy, T., 
Buttery, H.~J., Curran, J., Hunstead, R.~W., Piestrzynski, B., Robertson, 
J.~G., \& Sadler, E.~M.\ 2003, \mnras, 342, 1117 


\bibitem[Mehta et al.(2009)]{mehta09} Mehta, K.~T., 
Georganopoulos, M., Perlman, E.~S., Padgett, C.~A., 
\& Chartas, G.\ 2009, \apj, 690, 1706 

\bibitem[Meisenheimer et 
al.(1989)]{meisenheimer89} Meisenheimer, K., Roser, H.-J., Hiltner, P.~R., et al.\ 1989, \aap, 219, 63 

\bibitem[Mueller 
\& Schwartz(2009)]{mueller09} Mueller, M., \& Schwartz, D.~A.\ 2009, \apj, 693, 648 

\bibitem[Mullin 
\& Hardcastle(2009)]{mullin09} Mullin, L.~M., \& Hardcastle, M.~J.\ 2009, \mnras, 398, 1989 


\bibitem[Norman(1996)]{norman96} Norman, M.~L.\ 1996, Energy 
Transport in Radio Galaxies and Quasars, 100, 319 

\bibitem[Perlman et al.(2011)]{perlman11} Perlman, E.~S., et al.\ 
2011, arXiv:1107.2058 

\bibitem[Protassov et al.(2002)]{protassov02} Protassov, R., van 
Dyk, D.~A., Connors, A., Kashyap, V.~L., 
\& Siemiginowska, A.\ 2002, \apj, 571, 545 

\bibitem[Rawlings 
\& Saunders(1991)]{rawlings91} Rawlings, S., \& Saunders, R.\ 1991, \nat, 349, 138 

\bibitem[Sambruna et al.(2001)]{sambruna01} Sambruna, R.~M., Urry, 
C.~M., Tavecchio, F., Maraschi, L., Scarpa, R., Chartas, G., 
\& Muxlow, T.\ 2001, \apjl, 549, L161 

\bibitem[Sambruna et al.(2002)]{sambruna02} Sambruna, R.~M., 
Maraschi, L., Tavecchio, F., Urry, C.~M., Cheung, C.~C., Chartas, G., 
Scarpa, R., \& Gambill, J.~K.\ 2002, \apj, 571, 206 

\bibitem[Sambruna et al.(2004)]{sambruna04} Sambruna, R.~M., 
Gambill, J.~K., Maraschi, L., Tavecchio, F., Cerutti, R., Cheung, C.~C., 
Urry, C.~M., \& Chartas, G.\ 2004, \apj, 608, 698 

\bibitem[Scheuer(1995)]{scheuer95} Scheuer, P.~A.~G.\ 1995, 
\mnras, 277, 331 

\bibitem[Schlegel et al.(1998)]{schlegel98} Schlegel, D.~J., 
Finkbeiner, D.~P., \& Davis, M.\ 1998, \apj, 500, 525 

\bibitem[Schwartz et al.(2000)]{schwartz00} Schwartz, D.~A., et 
al.\ 2000, \apjl, 540, L69 

\bibitem[Schwartz(2002)]{schwartz02} Schwartz, D.~A.\ 2002, \apjl, 
569, L23 


\bibitem[Schwartz et al.(2006a)]{schwartz06a} Schwartz, D.~A., Marshall, H.~L., Lovell, J.~E.~J., et al.\ 2006a, \apj, 640, 592 

\bibitem[Schwartz et al.(2006b)]{schwartz06b} Schwartz, D.~A., Marshall, H.~L., Lovell, J.~E.~J., et al.\ 2006b, \apjl, 647, L107 


\bibitem[Schwartz et al.(2010)]{schwartz10} Schwartz, D.~A., 
Massaro, F., Siemiginowska, A., et al.\ 2010, International Journal of 
Modern Physics D, 19, 879 



\bibitem[Siemiginowska et al.(2002)]{siemiginowska02} Siemiginowska, 
A., Bechtold, J., Aldcroft, T.~L., Elvis, M., Harris, D.~E., 
\& Dobrzycki, A.\ 2002, \apj, 570, 543 


\bibitem[Siemiginowska et al.(2007)]{siemiginowska07} Siemiginowska, 
A., Stawarz, {\L}., Cheung, C.~C., Harris, D.~E., Sikora, M., Aldcroft, 
T.~L., \& Bechtold, J.\ 2007, \apj, 657, 145 

\bibitem[Sirianni et al.(2005)]{sirianni05} Sirianni, M., et al.\ 
2005, \pasp, 117, 1049 

\bibitem[Stawarz et al.(2004)]{stawarz04a} Stawarz, {\L}., Sikora, 
M., Ostrowski, M., \& Begelman, M.~C.\ 2004, \apj, 608, 95 

\bibitem[Suchkov \& Krist(1998)]{suchkov98} Suchkov, A., \& Krist, J.\ 1998, Instrument Science Report NICMOS 98 - 018, 11 pages, 18 

\bibitem[Tavecchio et 
al.(2003)]{tavecchio03} Tavecchio, F., Ghisellini, G., \& Celotti, A.\ 2003, \aap, 403, 83 

\bibitem[Tavecchio et al.(2000)]{tavecchio00} Tavecchio, F., 
Maraschi, L., Sambruna, R.~M., \& Urry, C.~M.\ 2000, \apjl, 544, L23 


\bibitem[Walker(1995)]{walker95} Walker, R.~C.\ 1995, Very Long 
Baseline Interferometry and the VLBA, 82, 133 

\bibitem[Wardle 
\& Aaron(1997)]{wardle97} Wardle, J.~F.~C., \& Aaron, S.~E.\ 1997, \mnras, 286, 425 

\bibitem[Willott et al.(1999)]{willott99} Willott, C.~J., 
Rawlings, S., Blundell, K.~M., \& Lacy, M.\ 1999, \mnras, 309, 1017 

\bibitem[Wilson et al.(2000)]{wilson00} Wilson, A.~S., Young, 
A.~J., \& Shopbell, P.~L.\ 2000, \apjl, 544, L27 

\bibitem[Wilson et al.(2006)]{wilson06} Wilson, A.~S., Smith, 
D.~A., \& Young, A.~J.\ 2006, \apjl, 644, L9 

\bibitem[Worrall 
\& Birkinshaw(2006)]{worrall06} Worrall, D.~M., \& Birkinshaw, M.\ 2006, Lecture Notes in Physics, 693, 39


\bibitem[Worrall(2009)]{worrall09} Worrall, D.~M.\ 2009, A\&ARv, 17, 1 


\bibitem[Wright 
\& Otrupcek(1990)]{wright90} Wright, A., \& Otrupcek, R.\ 1990, PKS Catalog (1990), 0 

\end{thebibliography}
\end{document}